\documentclass[aps,prl,reprint,groupedaddress]{revtex4-1}

\usepackage{color}
\usepackage{mathtools}
\usepackage{tikz}
\usepackage{enumitem}

\usepackage[hyperindex,breaklinks]{hyperref}
\usepackage{xcolor}
\usepackage[caption=false]{subfig}
\hypersetup{
    colorlinks=true,
    linkcolor=blue,
    citecolor=blue,
    urlcolor=blue
}
\usepackage{mathrsfs}

\usepackage{filecontents}

\begin{document}

\title{Compound Metaoptics for Amplitude and Phase Control of Wavefronts}
\author{Brian O. Raeker and Anthony Grbic}
\thanks{}
\affiliation{Department of Electrical Engineering and Computer Science, University of Michigan, Ann Arbor, MI 48109-2122 USA}

\date{\today}

\begin{abstract}
Metasurfaces allow tailored control over electromagnetic wavefronts. However, due to the local conservation of power flow, a passive, lossless, and reflectionless metasurface is limited to imparting phase discontinuities -- and not power density discontinuities -- onto a wavefront. Here, we show how the phase and power density profiles of a wavefront can be independently controlled using two closely-spaced, phase-discontinuous metasurfaces. The two metasurfaces, each designed to exhibit specific refractive properties, are separated by a wavelength-scale distance and together form a compound metaoptic. A systematic design procedure is presented, which allows transformation between arbitrary complex-valued field distributions without reflection, absorption, or active components. Such compound metaoptics may find applications in optical trapping of particles, displaying three-dimensional holographic images, shrinking the size of optical systems, or producing custom (shaped and steered) far-field radiation patterns.
\end{abstract}

\maketitle

Metasurfaces are two-dimensional arrays of sub-wavelength polarizable inclusions, which aggregrately manipulate an electromagnetic wave \cite{metasurface_rev1, metasurface_rev2, metasurface_rev3}. These inclusions (subwavelength-sized unit cells) are arranged in single- or few-layer stacks that are electrically/optically thin. In general, the electromagnetic interactions of a metasurface can be approximated as occuring at a boundary: simplifying analysis, design, and fabrication. A distinct property of metasurfaces is their ability to impart tailored phase discontinuities onto incident wavefronts. Utilizing this ability, metasurfaces have demonstrated optical functions such as focusing, refraction, and polarization control \cite{ yu_flat_optics_rev, shalaev_planar_photonics, pfeiffer_bianisotropy}. 

If a metasurface is restricted to be passive, lossless, and reflectionless, the power density profile of an incident wavefront is maintained when transmitted through the metasurface. Such metasurfaces exhibit high efficiency, but are limited to reshaping the phase profile of an incident wavefront and not its power density profile \cite{pfeiffer_huygens_ms, wong_wide_angle_ref_ms, chong_complex_wavefront_control}. As a result, a single phase-only metasurface is incapable of independently controlling both the phase and amplitude distributions of the transmitted field.  Specifically, this results in speckle noise (random fluctuations in amplitude across an image) in holographic images formed with a single phase-only surface \cite{ huang_3D_hologram}. Amplitude and phase control over an incident wavefront can suppress speckle, as shown by complex-valued holograms \cite{ shalaev_metasurface_hologram,speckle_reduction,speckle_reduction2}. However, such field control has not been demonstrated using completely transmissive metasurfaces.

\begin{figure}
 \centering
 \includegraphics[height = 1.65in]{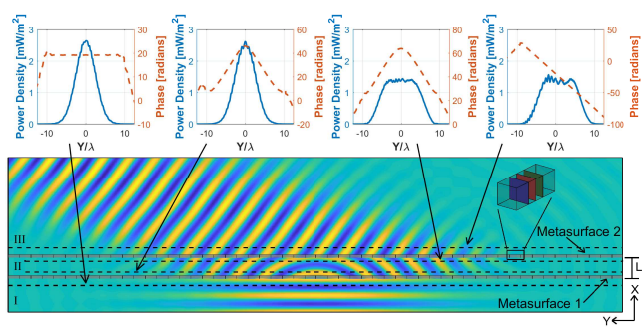}
 \caption{Two metasurfaces form the compound metaoptic, establishing three regions. The phase-discontinuous metasurfaces reshape the power density and phase profiles of an incident beam, as demonstrated by the displayed wavefront behavior. The inset plots display the power density (solid line) and phase (dashed line) profiles before and after each metasurface. A unit cell of the metasurface is also shown.
 \label{fig:sys_setup}}
 \end{figure}

Here, we introduce compound metaoptics which control the amplitude and phase of a wavefront in a passive, lossless, and reflectionless manner. A compound metaoptic is a collection of  individual metasurfaces arranged along an axis, analogous to an optical compound lens. With the additional degrees of freedom, compound metaoptics can achieve electromagnetic responses difficult or impossible to achieve with a single metasurface. We propose using paired, reflectionless metasurfaces, illustrated in Fig. \ref{fig:sys_setup}, to achieve both phase control (beam steering) and amplitude control (beam shaping) in a low-loss, low-profile manner.  This approach promises higher diffraction efficiencies than conventional holograms since both amplitude and phase are controlled with sub-wavelength pixelation.

Different methods of controlling the phase and amplitude of electromagnetic fields using metamaterials and metasurfaces have been reported. In \cite{gok}, control over phase and power flow within a reflectionless metamaterial is used to transform an incident wave in both amplitude and phase. This approach, however, requires an anisotropic and inhomogeneous medium through which the wavefront is reshaped. Additionally, leaky-wave structures \cite{tierney_LWA} and partially reflecting cavities \cite{epstein_arb_to_dir, raeker_cyl_ms, raeker_sph_ms}, which exploit reflections to establish the desired field,  also create complex valued aperture fields. However, these approaches either cannot be used with arbitrary sources or exhibit reflections impacting the source.

Metasurfaces have been proposed to control amplitude distributions by employing transmission loss  \cite{kim_tretyakov_AP_control}. These metasurfaces, however, trade efficiency for amplitude control. In contrast to these earlier works, we propose passive, lossless, and reflectionless compound metaoptics for arbitrary wavefront reshaping in terms of both amplitude and phase.

A pair of phase-discontinuous metasurfaces are used to mold the incident wavefront and form prescribed power density and phase distributions. The metasurfaces act as two phase planes: two reflectionless, inhomogeneous surfaces that locally manipulate the phase of the transmitted wavefront. Together, the two phase planes provide two degrees of freedom to control two wavefront characteristics: the power density and phase profiles. In the proposed arrangement, the first metasurface projects the desired amplitude onto the second metasurface. The second metasurface provides a phase-correction to realize the desired phase distribution. Overall, the pair of metasurfaces produce the desired complex-valued transmitted field.  The method is scalable from microwave to visible wavelengths. 

Related methods of forming optical fields with desired amplitude and phase distributions have used reflective spatial light modulators \cite{SLM_holo} or deformable mirrors \cite{Wu_deformed_mirrors}. The spatial light modulators or mirrors are located at conjugate Fourier planes of a two-lens optical system, limiting its compactness. Even lensless systems are still large due to reflective components \cite{holo_2SLM}. Due to the custom phase-discontinuity profiles implemented by the metasurfaces, the need for lenses and reflective components is avoided. This provides a significantly more low-profile solution to complex-valued field control. It allows the overall thickness of the metaoptic to be reduced to the scale of a few wavelengths. 

Figure \ref{fig:sys_setup} shows the layout of the compound metaoptic, where two phase-discontinuous metasurfaces are separated by a distance $L$. Huygens' metasurfaces locally control the transmission phase and can eliminate reflections by maintaining a wave impedance matched to the surrounding medium \cite{pfeiffer_huygens_ms}. Therefore, Huygens' metasurfaces are excellent candidates to provide the desired phase-discontinuous boundaries. Such metasurfaces can be implemented as multi-sheet structures, where the overall thickness is sub-wavelength (electrically/optically thin).

Extreme field control is required when transforming the amplitude and phase distributions of the source $E_{src}$ (incident on metasurface 1) to the desired complex-valued field $E_{des}$ (transmitted by metasurface 2) over a wavelength-scale distance $L$. Specifically, it requires wide angles of refraction at the two phase planes.  Huygens' metasurfaces, with induced electric and magnetic polarization currents, are practically reflectionless over a moderate range of incident/transmission angles. However, Huygens metasurfaces exhibit reflection for wide angles of refraction. These reflections are due to different wave impedances of the incident and transmitted fields. 

Such reflections can be mitigated using bianisotropic surface parameters: electric, magnetic, and magneto-electric responses. In addition to providing the transmission phase, bianisotropic metasurfaces serve as impedance matching layers. This  allows a reflectionless transition between a wave incident at one angle and refracted to another \cite{wong_wide_angle_ref_ms}. It should be noted that where wide-angle refraction is not required (e.g. for larger distances $L$, or when wave propagation is predominately paraxial), simple Huygens’ metasurfaces without a magneto-electric parameter may suffice.   

The design of the compound metaoptic involves three general steps. First, the field solution in region II (see Fig. \ref{fig:sys_setup}) is computed. This solution provides the phase-discontinuities to transform the incident field into the desired transmitted field. The second step is to compute the electromagnetic parameters of each metasurface, which are needed to realize the phase discontinuities. Finally, the metasurfaces are implemented as asymmetric cascades of electric surface impedance sheets \cite{wong_wide_angle_ref_ms, pfeiffer_bianisotropy}. 

A transverse electric field polarization with respect to the metasurface ($\hat z$-polarized) is assumed in this discussion, but the method also applies to the transverse magnetic polarization. To simplify the discussion, it is assumed the fields are invariant in the $z$-direction. In the two dimensional problems considered here, each metasurface is inhomogeneous along the $y$-direction and invariant in the $z$-direction. A time convention of $e^{i\omega t}$ is assumed.

With the source and desired (overall incident and transmitted) field profiles defined, the metasurface phase discontinuities are computed using a phase retrieval algorithm. Here, we use the Gerchberg-Saxton algorithm \cite{gerchberg_saxton, fineup_gs_exp}, modified for wave propagation in the near field. This algorithm numerically propagates the complex-valued field distributions between the two metasurfaces. The calculated field amplitude is replaced with the desired amplitude profile at each plane. The algorithm iterates until arriving at a field solution in region II, which links the two amplitude patterns via propagation. Ultimately, the algorithm results in a phase discontinuity at the first metasurface, $\phi_{MS1}$, such that
\begin{equation}
|E^z_{des}| = | \mathscr{F}_{MS2}^{-1} \{ \mathscr{F}_{MS1} \{ E^z_{src} \cdot e^{i\phi_{MS1}} \} \cdot e^{-ik_xL} \}|
\label{eq:amp_prop}
\end{equation}

Here $\mathscr{F}_{MS1}$ denotes a Fourier transform of the electric field at the first metasurface with respect to the $y$-coordinate, and $\mathscr{F}_{MS2}^{-1}$ denotes an inverse Fourier transform at the second metasurface over the plane wave spectrum $k_y$. The only variable parameter in eq. \eqref{eq:amp_prop} is $\phi_{MS1}$, thus one degree of freedom of the compound metaoptic is used to produce the desired amplitude pattern. With the distribution of $\phi_{MS1}$, the first metasurface projects the desired magnitude distribution onto the second metasurface. See the supplementary material for an in-depth discussion of this phase-retrieval algorithm \cite{supplemental_material}.

The phase discontinuity at the second metasurface is calculated as the phase difference between the adjacent fields. Overall, the manipulation of the source wavefront is determined by
\begin{equation}
E^z_{des} = \mathscr{F}_{MS2}^{-1} \{ \mathscr{F}_{MS1} \{ E^z_{src} \cdot e^{i\phi_{MS1}} \} \cdot e^{-ik_xL} \} \cdot e^{i\phi_{MS2}}.
\label{eq:E_match_calc}
\end{equation}

With $\phi_{MS1}$ determined using the phase-retrieval algorithm, $\phi_{MS2}$ is the only variable parameter in eq. \eqref{eq:E_match_calc}. Accordingly, $\phi_{MS2}$ is the second degree of freedom for the compound metaoptic, establishing the phase profile of the desired field. The two phase discontinuities are now defined, but local conservation of normal power flow must be enforced to ensure passive, lossless, and reflectionless metasurfaces.

Conserving normal power flow through each metasurface requires knowledge of the tangential electric and magnetic fields. Since the relative shapes of the electric field profiles are known from the phase-retrieval algorithm, the magnetic field plane wave spectrum $\mathscr{H}_y$ can be calculated. To do so, the electric field spectrum $\mathscr{E}_z$ is divided by the transverse electric wave impedance $(\eta_{_{TE}})$ for each plane wave component,
\begin{equation}
\mathscr{H}_y = \frac{\mathscr{E}_z}{\eta_{_{TE}}} = \frac{\mathscr{E}_z k_x}{\eta_0 k_0}
\end{equation}

The spatial magnetic field distribution is determined through an inverse Fourier transform of its plane wave spectrum. With the incident fields assumed known at the first metasurface, the transmitted fields are locally scaled to satisfy conservation of real power flow:
\begin{equation}
\text{Re}\{ E^i_z \times H^{i*}_y\}=\text{Re}\{E^t_z \times H_y^{t*}\}
\label{eq:real_power_flow}
\end{equation}
where $E^i_z$, $H^i_y$, $E^t_z$, and $H^t_y$ are the tangential incident and transmitted, electric and magnetic fields, respectively. The reflected fields are assumed to be zero. The plane wave spectrum of the field transmitted by the first metasurface is propagated forward to calculate the fields incident on the second metasurface. The conservation of power flow is enforced at the second metasurface.

With the field distributions fully determined throughout all three regions, the bianisotropic surface parameters of the metasurfaces are calculated. These parameters describe the surface properties implementing the conversions in wave impedance and phase \cite{pfeiffer_bianisotropy}. Since the field solutions in each region have been scaled to conserve power flow through the boundaries, these bianisotropic parameters model passive and lossless Huygens' surfaces. The surface parameters can be solved for in terms of the tangential fields, shown in the supplemental material \cite{supplemental_material} and similar to the approach in \cite{epstein_surf_params}.

\begin{figure}
 \centering
 \subfloat[\label{fig:huygens_surface_pic}]{
	 \includegraphics[width = 0.45\columnwidth]{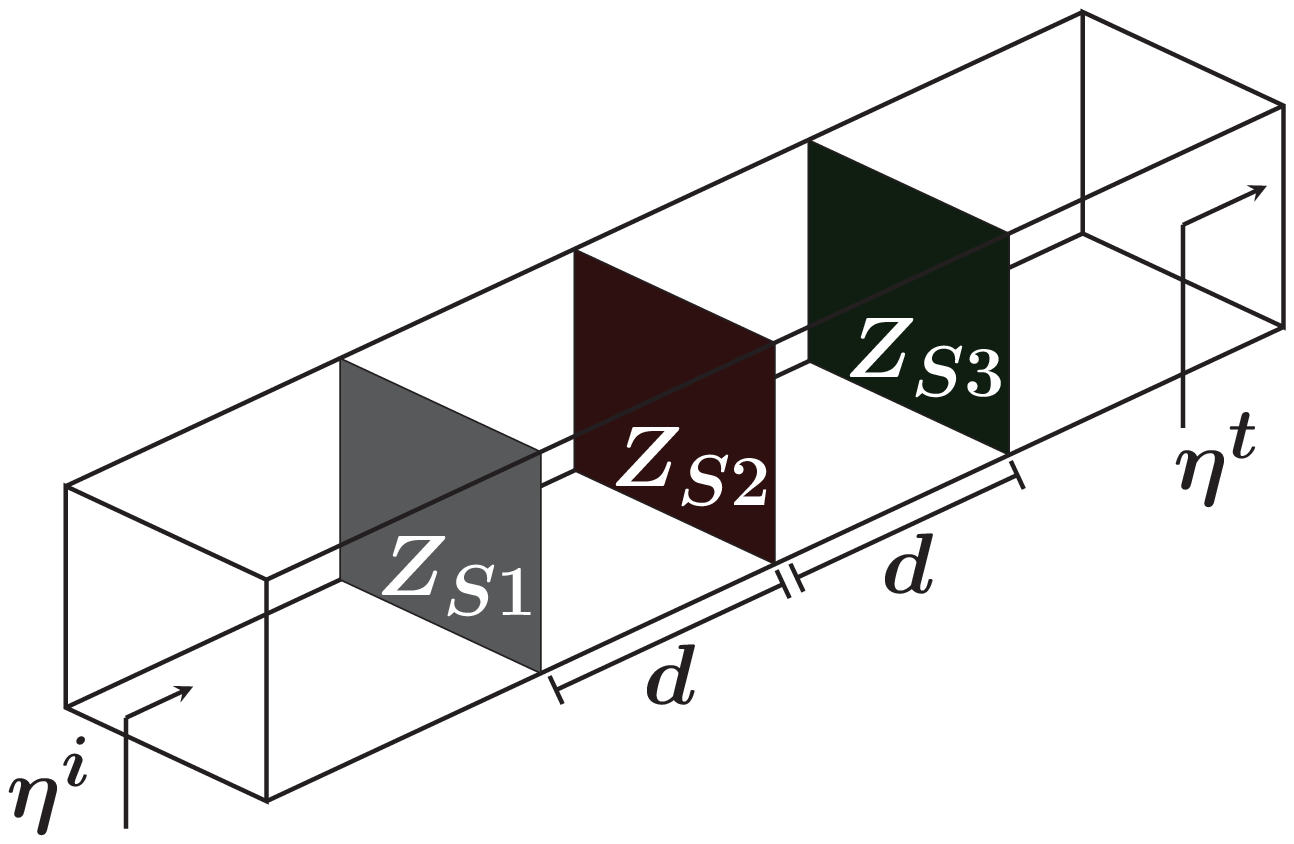}%
	}
	\subfloat[\label{fig:circuit_diagram}]{
	 \includegraphics[width=0.54\columnwidth]{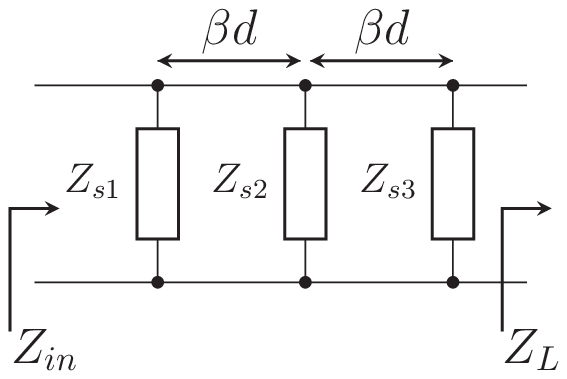}%
	}
 \caption{The unit cell of a bianisotropic Huygens' metasurface is shown in (a), where three sheet impedances $(Z_s)$ are separated by a distance $d$. The wave impedance on either side of the metasurface is denoted as $\eta^i$ for the incident field and $\eta^t$ for the transmitted field. The unit cell is modeled by the transmission line circuit shown in (b), where transmission lines separate three shunt impedances. The input and load impedances $(Z_{in}$ and $Z_L)$ are equal to the wave impedances. \label{fig:huygens_surface}}
 \end{figure}

The ideal performance of the Huygens' metasurfaces can be observed by explicitly defining the desired  electric and magnetic surface current densities in place of the metasurfaces. In accordance with the Surface Equivalence Principle, the fields in regions II and III are transformed as desired due to these current distributions. The wavefront behavior in Fig. \ref{fig:sys_setup} displays such a \emph{COMSOL Multiphysics} simulation where an incident Gaussian beam is expanded and refracted to 55 degrees. 

While this approach results in surface parameters that reshape the incident wavefront in amplitude and phase, these parameters must be translated into realizable metasurface designs. Here, we make use of bianisotropic Huygens' metasurfaces, which consist of a cascade of electric impedance sheets \cite{pfeiffer_bianisotropy, wong_wide_angle_ref_ms,Pfeiffer_light_bending}. Figure \ref{fig:huygens_surface_pic} shows a Huygens' metasurface unit cell, where three electric sheet impedances are separated by a sub-wavelength distance $d$. Unit cells of this structure can exhibit equivalent electric and magnetic current densities and be placed side-by-side to produce a gradient metasurface.
 
To analyze the metasurface unit cell of Fig. \ref{fig:huygens_surface_pic}, we can model it as a transmission-line circuit, shown in Fig. \ref{fig:circuit_diagram}. The transmission-line model contains three shunt impedance values (representing the impedance sheets), separated by an electrical length of $\beta d$. The input and load impedances $(Z_{in}$ and $Z_L)$ of the transmission-line model are taken to be the transverse electric wave impedances on either side of the metasurface.

The three variable parameters (shunt impedances) of the circuit model allow control over three desired characteristics of each unit cell. We choose these to be: (1) input impedance matched to the local incident wave impedance, (2) load impedance matched to the local transmitted wave impedance, and (3) a desired phase delay through the surface. Matching the input and load impedances eliminates reflections from the boundary and the desired phase delay implements the local metasurface phase discontinuity. Since the tangential fields are known adjacent to both metasurfaces, individual unit cell parameters are defined to locally satisfy these distributions. See the supplemental material for the derivation of the impedance sheet values as a function of the three desired characteristics \cite{supplemental_material}.

Using the procedure we have described, the compound metaoptic is designed such that an incident wavefront is altered in amplitude and phase to produce a desired complex field distribution. We provide two simulation examples where an incident Gaussian beam (beam radius of $5\lambda$) is manipulated using a compound metaoptic. 

\begin{figure}
\subfloat[\label{fig:ex2_Efield}]{
	\includegraphics[width=0.46\columnwidth]{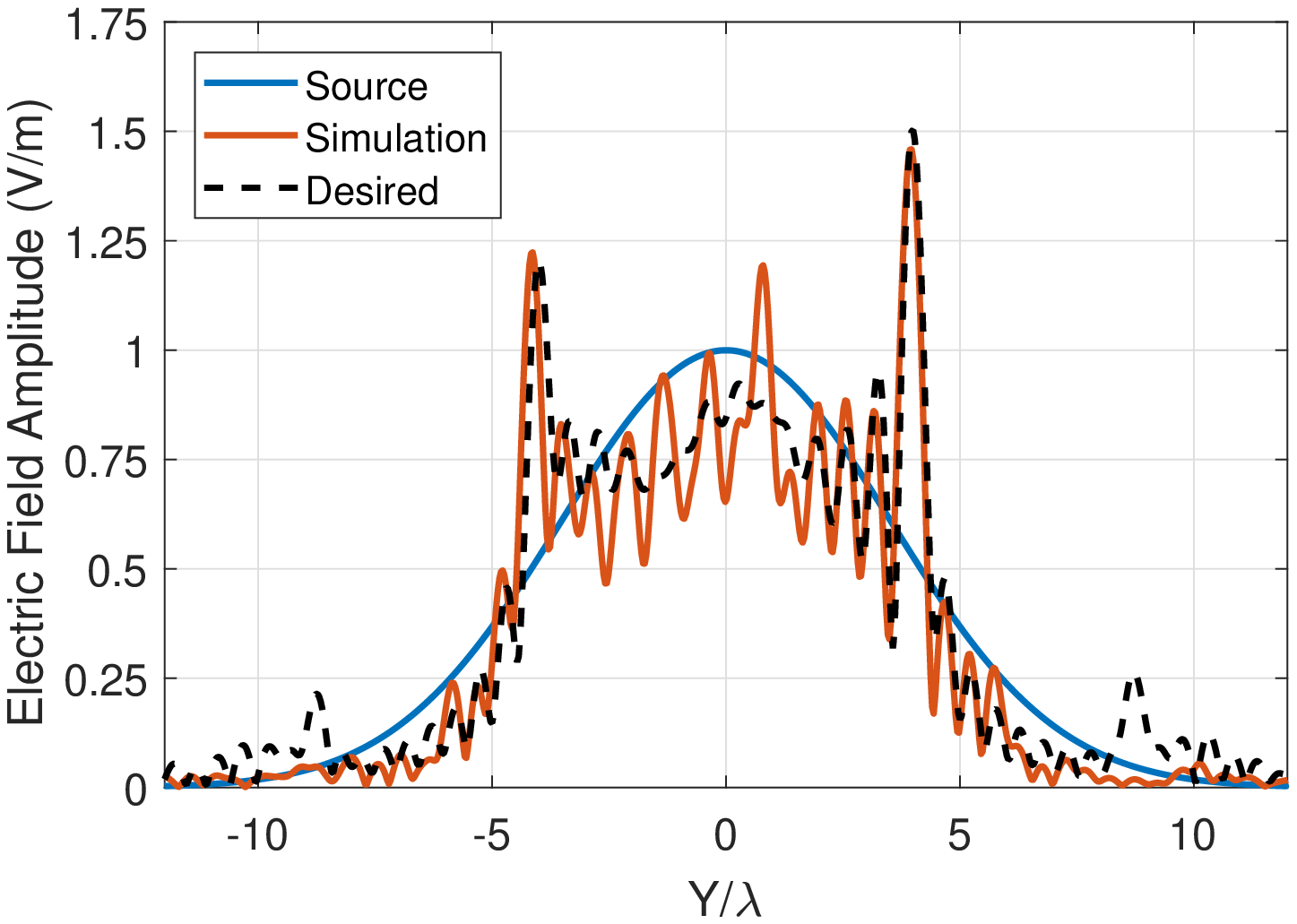}
	}
\subfloat[\label{fig:ex2_ffield}]{
	\includegraphics[width=0.46\columnwidth]{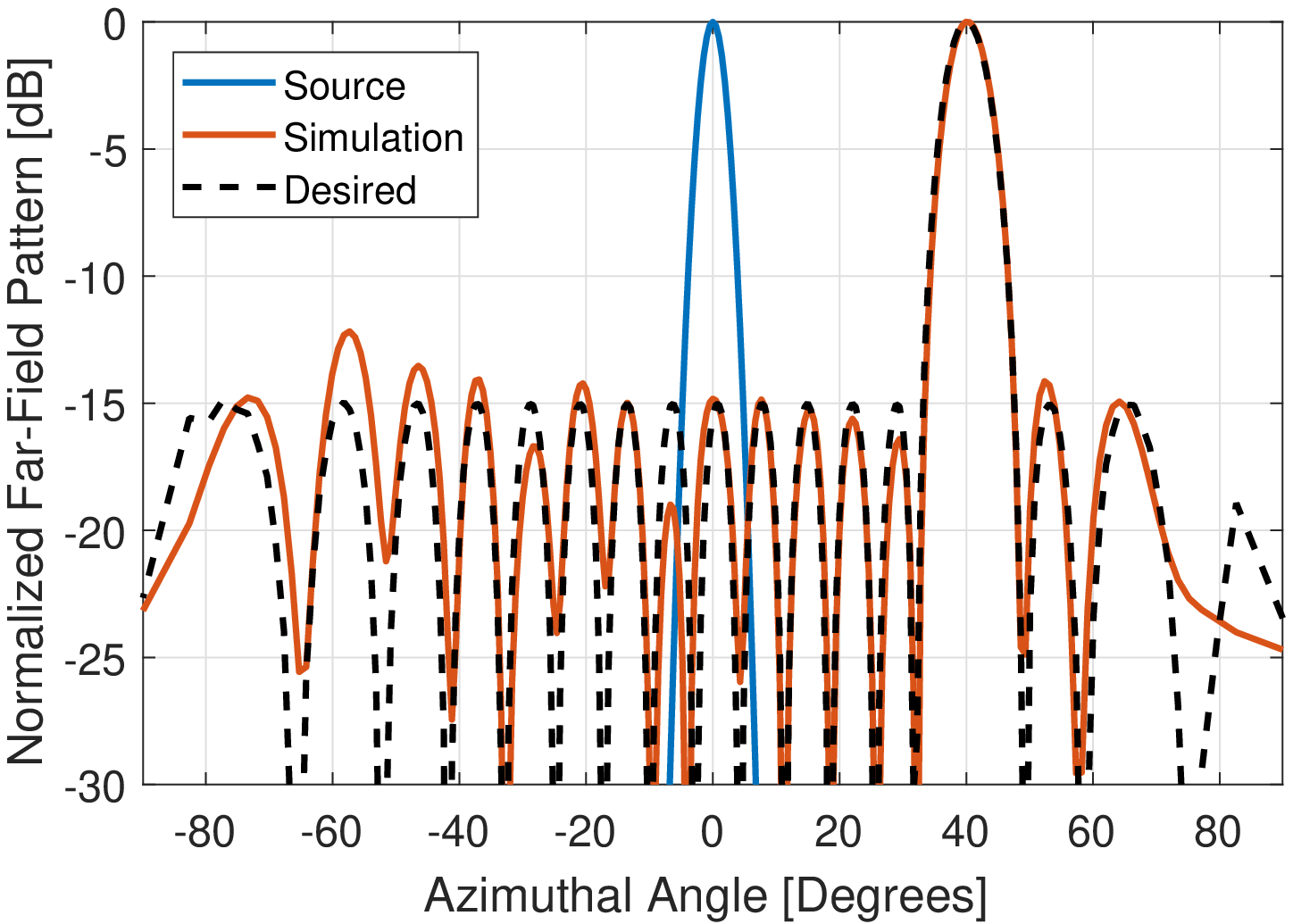}
	}\hfill
\subfloat[\label{fig:DTch_Efield}]{
	\includegraphics[width=0.95\columnwidth]{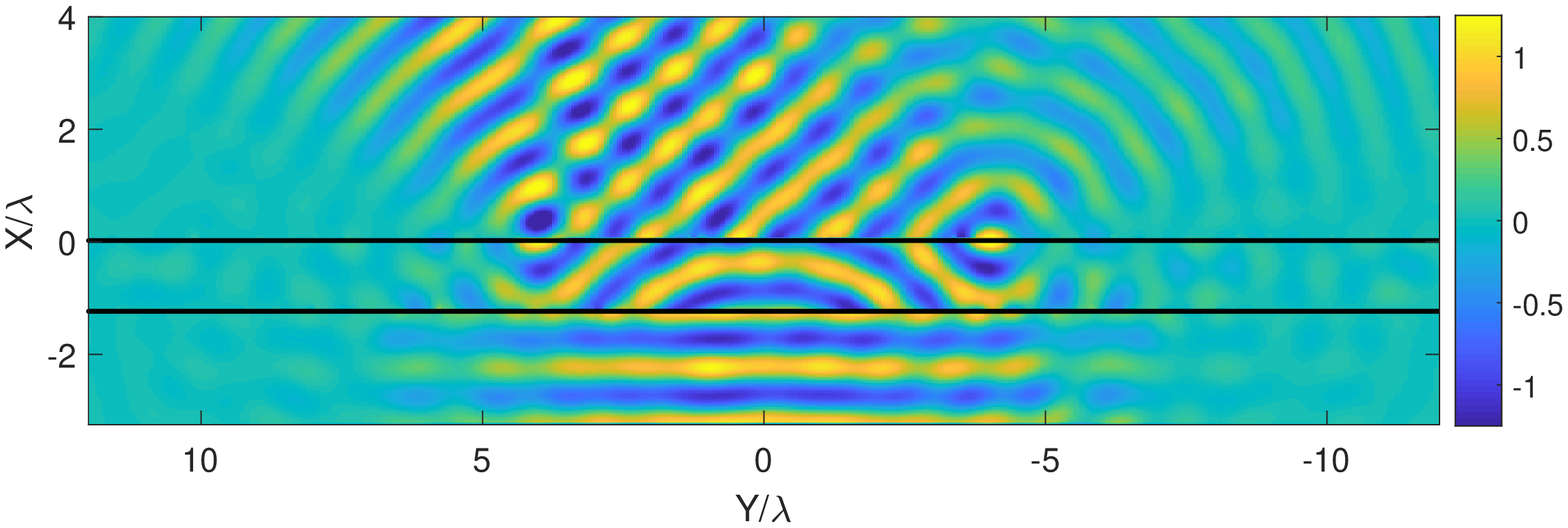}
	}
\caption{\label{fig:ex2_E_field} A compound metaoptic is designed to re-shape an incident Gaussian beam to produce a Dolph-Tchebyscheff far-field pattern pointed towards 40 degrees. The metaoptic performance is shown in (a) as the electric field amplitude, (b) as the far-field radiation pattern, and (c) as the real part of the simulated electric field.}
\end{figure}

In the first example, the amplitude and phase profiles of the incident Gaussian beam are re-shaped to produce a Dolph-Tchebyscheff far-field pattern pointed toward 40 degrees. This far-field pattern exhibits the narrowest main beam since all sidelobes are at the same level \cite{rad_systems}. Sinc function interpolation of the discrete array element weights was employed to determine an equivalent continuous electric field distribution. Figure \ref{fig:ex2_Efield} shows the desired amplitude distribution producing a far-field pattern having sidelobes of -15 dB.

The sheet impedance values of the metasurfaces were calculated for a separation distance of $L=1.25\lambda$, a unit cell width of $\lambda/16$, and an impedance sheet separation of $d=\lambda/80$. The sheet impedances were modeled as ideal impedance boundaries in the commercial full-wave electromagnetics solver \emph{COMSOL Multiphysics}. Figure \ref{fig:ex2_ffield} shows the far-field pattern of the metaoptic, which closely matches the desired Dolph-Tchebyscheff pattern. Each of the sidelobes are nearly -15 dB relative to the main lobe and all pattern nulls are located at the correct angle. Figure \ref{fig:DTch_Efield} shows the simulated electric field, where the first metasurface projects the desired amplitude distribution across the separation distance $L$ and the second metasurface points the main beam toward 40 degrees. Figure \ref{fig:DTch_Efield} also shows there are nearly no reflections from the compound metaoptic.

\begin{figure}
\subfloat[\label{fig:3ptsc_geom}]{
	\includegraphics[width=0.46\columnwidth]{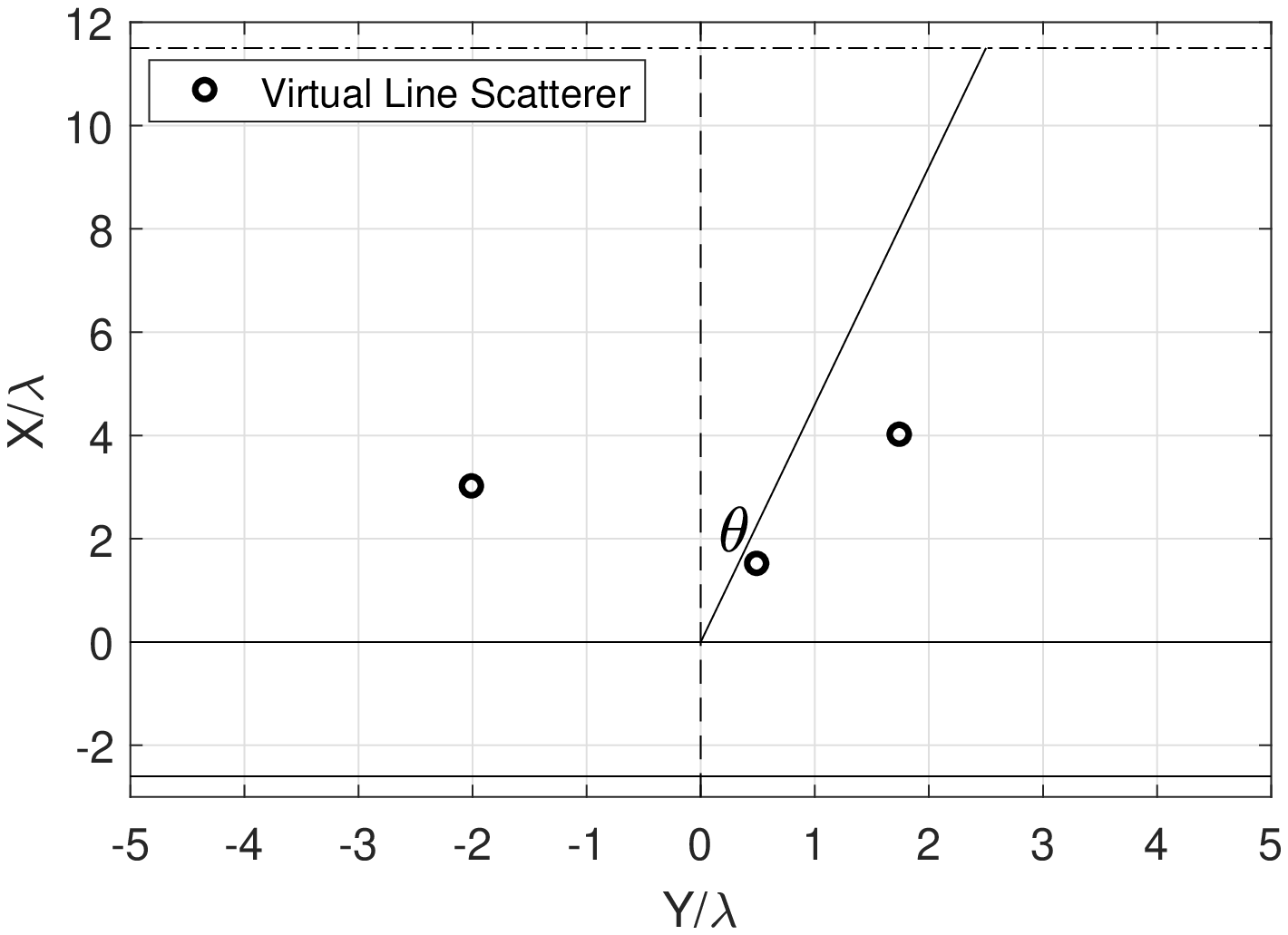}
	}
\subfloat[\label{fig:3ptsc_Eamp_nf}]{
	\includegraphics[width=0.46\columnwidth]{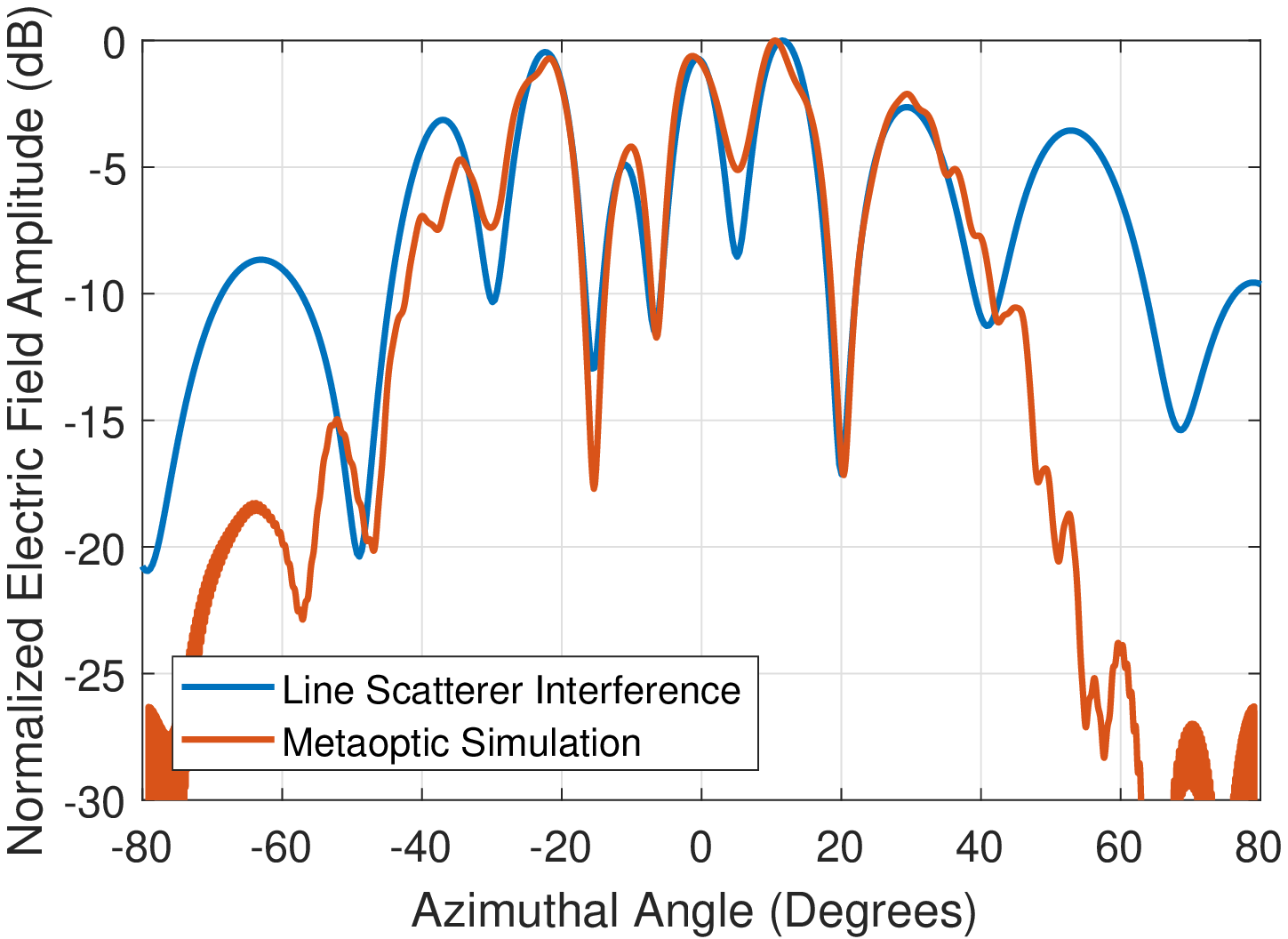}
	}\hfill
\subfloat[\label{fig:3ptsc_Ephase_nf}]{
	\includegraphics[width=0.46\columnwidth]{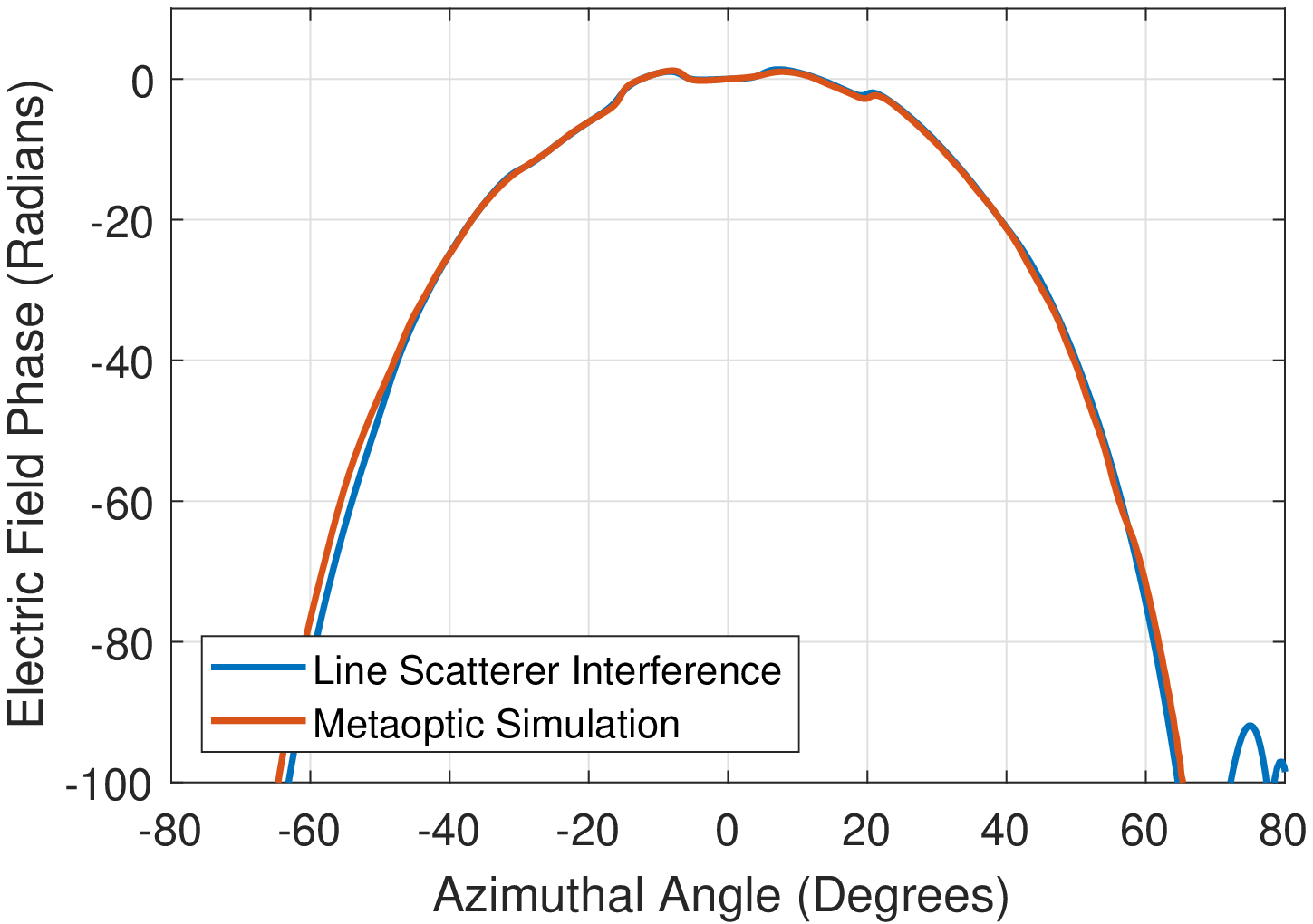}
	}
\subfloat[\label{fig:3ptsc_Eamp_ff}]{
	\includegraphics[width=0.46\columnwidth]{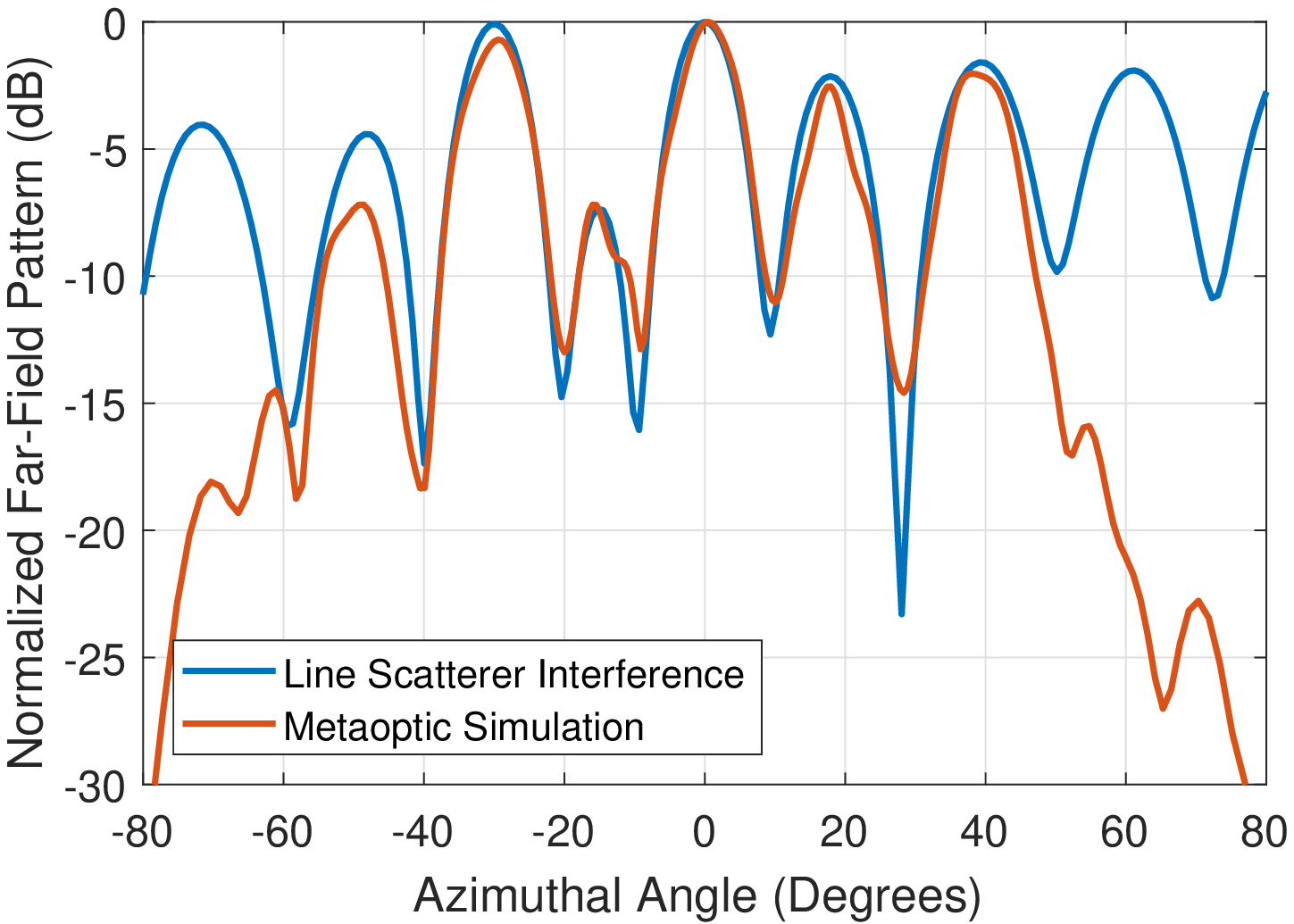}
	}
\caption{A compound metaoptic produces the field scattered by three line scatterers arranged as shown in (a). The simulated electric field is compared to the desired scattered field at a distance of $11.5\lambda$ from the metaoptic in (b) as the field amplitude and (c) as the phase. The far-field radiation pattern is shown in (d) for the simulated and desired scattered field distributions.\label{fig:3ptsc}}
\end{figure}

In the second example, a compound metaoptic is designed to radiate a field identical to the field scattered by three line scatterers. In other words, a simple complex-valued hologram of the scatterers is realized using the compound metaoptic. The virtual line scatterers are in the region beyond the metasurface system $(x>0)$, as shown in Fig. \ref{fig:3ptsc_geom}. The scattered plane wave spectrum associated with each scatterer is calculated for an incident plane wave traveling in the $-x$ direction. The spectra are summed together to obtain the plane wave spectrum of the total scattered field along the $x=0$ plane. A windowing function was applied to this spectrum such that the scattered field is visible over an azimuthal range between $\pm40$ degrees. The desired spatial electric field distribution is obtained from the windowed plane wave spectrum and used to design the compound metaoptic.  

The metaoptic was designed with a separation distance of $L=2.25\lambda$, a unit cell dimension of $\lambda/16$, and an impedance sheet spacing of $d=\lambda/60$. Figure \ref{fig:3ptsc_Eamp_nf} and Fig. \ref{fig:3ptsc_Ephase_nf} compare the simulated electric field amplitude and phase, respectively, at a distance of $11.5\lambda$ from the metaoptic with the desired interference pattern of the three line scatterers. We see that the electric field produced by the metaoptic closely matches, in amplitude and phase, the ideal interference pattern of the three line scatterers over the wide azimuthal range between $\pm40$ degrees. This is achieved even at short distances from the metaoptic. Figure \ref{fig:3ptsc_Eamp_ff} shows the far-field pattern also closely matches the true interference pattern over the desired azimuthal range. This demonstrates that the compound metaoptic is capable of reconstructing the field scattered from known objects in amplitude and phase.

The proposed compound metaoptic makes use of two phase-discontinuous metasurfaces to mold the available power density from the source field into a desired phase and power density distribution. The metaoptic can have a wavelength-scale thickness due to bianisotropic properties of the constitutive Huygens' metasurfaces. 

The proposed compound metaoptic concept may find applications in 3D holographic display technology.The approach also presents a new design paradigm for electronically scanned antennas.  Conventional approaches at microwave/millimeter-wave frequencies utilize a phased array, where phase shifters provide beam steering and amplifiers/attenuators provide beam shaping. Such a method becomes increasingly difficult to implement at shorter wavelengths due to transistor cutoff frequencies and array feeding network losses. The proposed approach is especially attractive for millimeter-wave frequencies and beyond, given that it allows beam shaping (amplitude control) and beam steering (phase control) simply by using two phase planes.

\begin{acknowledgments}
This work was supported by the National Science Foundation Graduate Research Fellowship Program under Grant No. DGE 1256260 and the Office of Naval Research under Grant No. N00014-15-1-2390. 
\end{acknowledgments}

\bibliography{main_raeker_compound_metaoptic}

\end{document}


\begin{center}
\textbf{\large Supplemental Material}
\end{center}

\section{Modified Gerchberg-Saxton phase retrieval algorithm}

In the main text, the phase discontinuity profiles for each metasurface of the compound metaoptic are computed using a modified version of the Gerchberg-Saxton phase retrieval algorithm. The Gerchberg-Saxton algorithm iteratively reconstructs the phase profile of a wavefront from two intensity measurements, taken at different planes \citep{gerchberg_saxton,fineup_gs_exp}. Since the field profiles are actually complex-valued at each plane, the two intensity profiles are partial constraints which that must be satisfied when determining the wavefront's phase profile. The wavefront phase profile is iteratively determined so that the intensity measurements are linked by propagation of the wavefront between the two planes. The metasurface phase discontinuities can be calculated as the difference between the wavefront phase profile, and the incident/desired field phase profile. 

The Gerchberg-Saxton algorithm most commonly reconstructs a wavefront's phase profile using intensity measurements taken at planes in the near field and the far field, or before and after a lens. In this case, a single Fourier transform is used to propagate the complex-valued field from the first plane to the second, and an inverse Fourier transform is used to reverse propagate the complex-valued field from the second plane to the first. 

However, since the compound metaoptic has a thickness wavelength in scale, the field profiles are propagated between planes that are both in the near field. We modified the propagation step of the Gerchberg-Saxton algorithm by replacing the single Fourier and inverse Fourier transforms with a plane wave propagation procedure. The procedure involves taking the Fourier transform of the complex-valued field in the spatial domain, propagating the plane wave spectrum to the other plane, and then taking the inverse Fourier transform to obtain the complex-valued field in the spatial domain at the other plane. The rest of the Gerchberg-Saxton algorithm is then continued. At each plane, the amplitude of the propagated field is replaced with the desired amplitude profile. With each iteration, the phase estimate of the wavefront is improved until the intensity measurements at two closely spaced planes match the desired amplitude profiles.

In the main text, the desired amplitude profiles are denoted as $|E_{src}|$ at the first plane (first metasurface) and $|E_{des}|$ at the second plane (second metasurface). The modified Gerchberg-Saxton algorithm estimates the complex-valued field at each plane ($E_{MS1}$ or $E_{MS2}$) by replacing the amplitude profile of the calculated field with the desired field amplitude profiles. In this way, the field estimate will satisfy the amplitude profile partial constraints. As the algorithm iterates, the phase profile of the field estimates improve so that the propagated field better matches the desired amplitude profile at the other plane. The phase of the field transmitted from the first plane (first metasurface) is denoted as $\phi_{MS1}^t$ and the phase of the field incident on the second plane (second metasurface) is denoted as $\phi_{MS2}^i$. The algorithm used in the main text follows four general steps. For a single iteration of the algorithm:
\begin{enumerate}
	\item An estimate of the complex field at the first plane $(E_{MS1})$ is formed by multiplying the amplitude of the source electric field $(|E_{src}|)$ by the current estimate of the wavefront phase profile $(\phi^t_{MS1})$. This field estimate satisfies the amplitude profile partial constraint required at plane 1. For the first iteration, any phase profile for $\phi^t_{MS1}$ can be provided since the algorithm will converge on a solution.
	\item The field estimate at the first plane is propagated to the second plane using the plane wave propagation procedure. Propagation of the plane wave spectrum amounts to multiplying each spectral component by the phase term $e^{-ik_xL}$. The resulting complex-valued field profile may not accurately match the desired field amplitude ($|E_{des}|$) at the second plane, but improves as the algorithm iterates.
	\item The magnitude of the propagated complex field is replaced with the desired amplitude profile, $|E_{des}|$, to form an estimate of the complex field at the second plane ($E_{MS2}$). This action applies the current estimate of the wavefront's phase profile $(\phi_{MS2}^i)$ to the desired amplitude profile $(|E_{des}|)$.
	\item This field estimate is reverse propagated to the first plane using the plane wave propagation procedure. The phase term $e^{ik_xL}$ is used to account for reverse propagation. The phase of this calculated field becomes the current estimate of the wavefront phase profile ($\phi^t_{MS1}$) and is used in the next iteration of the algorithm. 
\end{enumerate}

\begin{figure}[!ht]
 \centering
 \includegraphics[height = 2in]{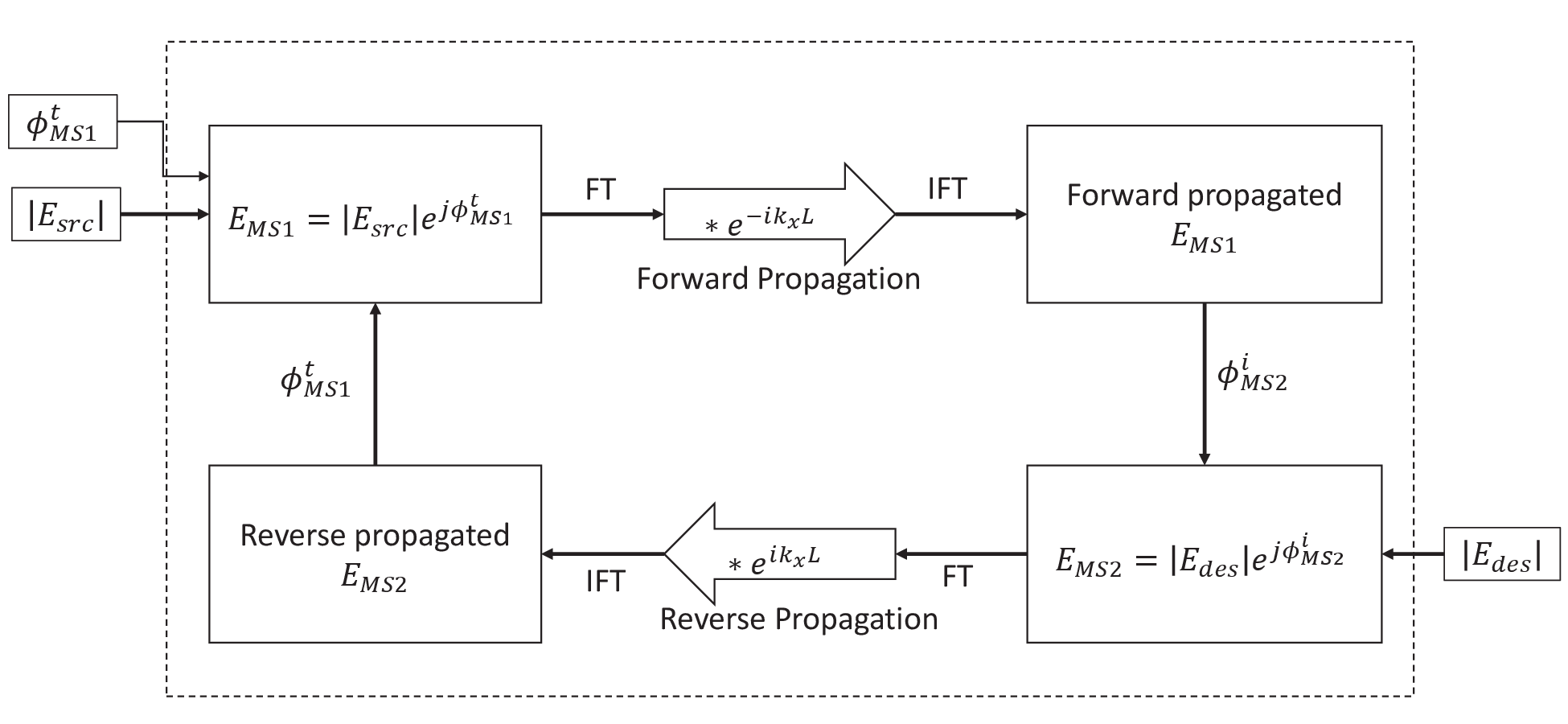}
 \caption{The modified Gerchberg-Saxton algorithm used to iteratively determine a phase profile which links the two amplitude measurements by propagation. Here, $E_{MS}$ is the field estimate at metasurface 1 or 2, $E_{src}$ is the source field, and $E_{des}$ is the desired field pattern. An (inverse) Fourier transform in the $y$-direction is denoted as (I)FT.\label{fig:GS_alg}}
 \end{figure}

These steps are repeated until the error between the propagated field distributions and the desired amplitude profile at that plane is acceptable. Once the algorithm converges, the phase distribution of the fields within region II are known. Fig. \ref{fig:GS_alg} shows a flowchart of the modified Gerchberg-Saxton algorithm.

Once the modified Gerchberg-Saxton algorithm results in the desired phase profile of the wavefront, the phase discontinuity profiles of each metasurface ($\phi_{MS1}$ and $\phi_{MS2}$) are determined by calculating the difference in phase of the tangential fields.

\begin{align}
\phi_{MS1} =& \phi_{MS1}^t - \measuredangle E_{src}\\	
\phi_{MS2} = & \measuredangle E_{des} - \phi_{MS2}^i
\end{align}

\section{Derivation of Surface Parameters in terms of Tangential Fields}

A surface boundary between two regions can be described in terms of an electric admittance $(Y)$, magnetic impedance $(Z)$, and magneto-electric terms $(\chi, \Upsilon)$ \cite{metasurface_rev1, idemen_disc_fields}. These surface parameters relate the averaged tangential fields ($\bf{E}_{avg}$ and $\bf{H}_{avg}$) on either side of the boundary to the electric and magnetic current densities ($\bf J$ and $\bf M$) induced on the boundary:
\begin{equation}
\begin{bmatrix}
\bf{J}\\
\bf{M}
 \end{bmatrix}
 =
 \begin{bmatrix}
 \overline{\overline{Y}} & \overline{\overline{ \chi}} \\
 \overline{\overline{ \Upsilon }}& \overline{\overline{ Z}}
 \end{bmatrix}
 \begin{bmatrix}
\bf{E_{avg}}\\
\bf{H_{avg}}
 \end{bmatrix}
\end{equation}

These relations can be manipulated such that the surface parameters are expressed in terms of the incident and transmitted fields, similar to the procedure in \cite{epstein_surf_params}. Given a TE incident polarization $({\bf E}=E_z \hat{z}$, ${\bf H}=H_y \hat{y})$ and assuming a reciprocal surface, the surface parameters can be simplified to
\begin{align}
\overline{\overline{Y}} = Y\begin{bmatrix}
1 & 0\\0 & 1
\end{bmatrix}
\qquad
\overline{\overline{Z}} = Z\begin{bmatrix}
1 & 0\\0 & 1
\end{bmatrix}
\qquad
\overline{\overline{\chi}} = \chi\begin{bmatrix}
0 & 1\\ -1 & 0
\end{bmatrix}
\qquad
\overline{\overline{\Upsilon}} = -\overline{\overline{\chi^T}}= \chi\begin{bmatrix}
0 & 1\\ -1 & 0
\end{bmatrix}
\end{align}

With the metasurface in the $YZ$ plane, the induced surface currents then become:
\begin{equation}
J_z = Y E_{z,avg}-\chi H_{y,avg}
\qquad
M_y = -\chi E_{z,avg} - Z H_{y,avg}.
\end{equation}

Here $E_z^i$ and $H_y^i$ are the tangential electric and magnetic field components of the incident wavefront, and $E_z^t$ and $H_y^t$ are the tangential components of the transmitted fields. After applying the boundary conditions, and assuming no reflected field, we have two complex equations with three complex unknowns
\begin{align}
\frac{1}{2}Y(E_z^t -E_z^i) &= H_y^t\left(1-\frac{\chi}{2}\right) - H_y^i\left(1+\frac{\chi}{2}\right)\\
E_z^t-E_z^i &= -\frac{\chi}{2}(E_z^t + E_z^i) - \frac{Z}{2}(H_y^t + H_y^i)
\end{align}

Conservation of normal power density requires 
\begin{equation}
\text{Re}\left\{E_z^i H_y^{i*}\right \} = \text{Re}\left\{E_z^t H_y^{t*}\right \},
\end{equation}
and the lossless condition mandates that 
\begin{equation}
\text{Re}\{ Y \} = \text{Re}\{ Z \} = \text{Im}\{ \chi \}=0.
\end{equation}

Using all of these conditions, and solving for the surface parameters, results in 
\begin{align}
\text{Re}\{ \chi \} &= - \frac{2 \text{Re} \{ E_z^t H_y^{i*} - E_z^{i} H_y^{t*} \}}{\text{Re} \{ (E_z^i + E_z^t)(H_y^i + H_y^t)^* \}}\\
\text{Im}\{ Y \}&= \frac{2\text{Im}\{ (E_z^i + E_z^t)^* (H_y^i - H_y^t)\} + \text{Re}\{ \chi \} \text{Im}\{ (E_z^i + E_z^t)^* (H_y^i + H_y^t)\}}{|E_z^i + E_z^t|^2}\\
\text{Im}\{ Z \}&= -\frac{2\text{Im}\{ (E_z^t - E_z^i) (H_y^i + H_y^t)^*\} - \text{Re}\{ \chi \} \text{Im}\{ (E_z^i + E_z^t) (H_y^i + H_y^t)^*\}}{|H_y^i + H_y^t|^2}
\end{align}
where the surface parameters are now defined in terms of the tangential fields. When a surface with these parameters is illuminated by $E_z^i$ and $H_y^i$, then $E_z^t$ and $H_y^t$ will be transmitted with lossless and reflectionless performance. It should be noted that these equations are only valid when the field distributions locally satisfy conservation of normal power flow.

\section{\label{sec:sheet_impedance} Derivation of the Surface Impedance Values for a Huygens' Metasurface Unit Cell} 

The impedance sheet values for each unit cell of the bianisotropic Huygens' metasurfaces can be derived from three quantities: the wave impedance (ratio of tangential electric and magnetic fields) on either side of the metasurface and the desired phase discontinuity. To do so, we first derive the necessary overall impedance matrix of the unit cell, and then convert the impedance matrix to an ABCD matrix. The ABCD matrix is then used to derive expressions for each of the impedance sheets.

Each unit cell of the metasurface can be described as a two port network having an impedance matrix,
\begin{equation}
{\bf Z} = \begin{bmatrix}
Z_{11} & Z_{12} \\
Z_{21} & Z_{22}
\end{bmatrix}=
j \begin{bmatrix}
X_{11} & X_{12} \\
X_{21} & X_{22}
\end{bmatrix}
\end{equation}
where port 1 is considered the input port and port 2 the load port. In addition, the impedance matrix must be fully imaginary if the unit cell is lossless.

\begin{figure}[h]
\centering
 \includegraphics[height = 1.5in]{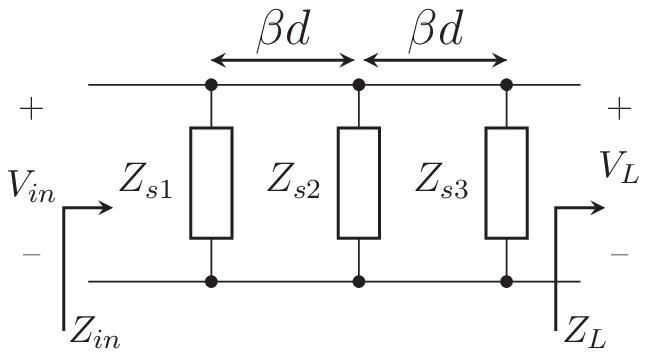}
 \caption{The unit cell of the Huygens' metasurface labeled with the corresponding voltage and impedance quantities.\label{fig:trl_circuit}}
\end{figure}

Given the lossless condition, real power flow normal to the surface must be conserved across each unit cell. The load voltage $(V_L)$ can be expressed in terms of the input voltage $(V_{in})$, the input impedance $(Z_{in} =R_{in} + jX_{in} =|Z_{in}|e^{j\phi_{Zin}})$, and the load impedance $(Z_{L} =R_{L} + jX_{L}=|Z_{L}|e^{j\phi_{ZL}})$ as given by \ref{eq:relate_v}. Figure \ref{fig:trl_circuit} shows the relation of these values to the unit cell diagram.

\begin{align}
\text{Re}\{V_{in} I_{in}^*\} & = \text{Re}\{V_L I_L^*\} \nonumber \\
\text{Re}\left\{V_{in} \left(\frac{V_{in}}{Z_{in}}\right)^*\right\} & = \text{Re}\left\{V_L \left(\frac{V_L}{Z_{L}}\right)^*\right\}\nonumber \\
\frac{|V_{in}|^2 R_{in}}{R_{in}^2 + X_{in}^2} &= \frac{|V_L|^2 R_{L}}{R_{L}^2 + X_{L}^2}\nonumber\\
|V_L|&=|V_{in}|\sqrt{\frac{R_{in}}{R_{L}} \frac{R_{L}^2 + X_{L}^2}{R_{in}^2 + X_{in}^2}}\nonumber \\
|V_L| &= |V_{in}|R
\label{eq:relate_v}
\end{align}

The desired phase shift across the unit cell is defined as $\phi = \phi_{V_L}-\phi_{V_{in}}$. The voltages and currents flowing into the input and load ports of the unit cell are related by the impedance matrix
\begin{equation}
\begin{bmatrix}
V_{in} \\V_L
\end{bmatrix} = 
j \begin{bmatrix}
X_{11} & X_{12} \\
X_{21} & X_{22}
\end{bmatrix}
\begin{bmatrix}
I_{in}\\I_L
\end{bmatrix}
\end{equation}

By expressing $V_L$, $I_{in}$, and $I_L$ in terms of $V_{in}$, $Z_{in}$, and $Z_L$, we have 

\begin{align}
\begin{bmatrix}
V_{in} \\|V_L|e^{j\phi_{V_L}}
\end{bmatrix} &= 
j \begin{bmatrix}
X_{11} & X_{12} \\
X_{21} & X_{22}
\end{bmatrix}
\begin{bmatrix}
\frac{V_{in}}{Z_{in}}\\-\frac{|V_L|}{Z_L}e^{j\phi_{V_L}}
\end{bmatrix} \\
\begin{bmatrix}
V_{in} \\|V_{in}|Re^{j\phi_{V_L}}
\end{bmatrix} &= 
j \begin{bmatrix}
X_{11} & X_{12} \\
X_{21} & X_{22}
\end{bmatrix}
\begin{bmatrix}
\frac{V_{in}}{Z_{in}}\\-|V_{in}|\frac{R}{Z_L}e^{j\phi_{V_L}}
\end{bmatrix} \\
\begin{bmatrix}
V_{in} \\V_{in} Re^{-j\phi_{V_{in}}}e^{j\phi_{V_L}}
\end{bmatrix} &= 
j \begin{bmatrix}
X_{11} & X_{12} \\
X_{21} & X_{22}
\end{bmatrix}
\begin{bmatrix}
\frac{V_{in}}{Z_{in}}\\-V_{in}\frac{R}{Z_L}e^{-j\phi_{V_{in}}}e^{j\phi_{V_2}}
\end{bmatrix} \\
\begin{bmatrix}
1 \\Re^{j\phi}
\end{bmatrix} &= 
j \begin{bmatrix}
X_{11} & X_{12} \\
X_{21} & X_{22}
\end{bmatrix}
\begin{bmatrix}
\frac{1}{Z_{in}}\\-\frac{R}{Z_L}e^{j\phi}
\end{bmatrix} 
\end{align}

This matrix equation provides four separate equations by equating the real and imaginary terms. The four entries of the impedance matrix are solved to be in terms of the input and load impedances and the delay across the unit cell.

\begin{align}
X_{11} &= X_{12} \frac{R|Z_{in}|}{|Z_{L}|} \frac{\cos(\phi_{Z_L}-\phi)}{\cos(\phi_{Zin})}\\
X_{12} &= \frac{|Z_{L}|}{R\left[\cos(\phi_{Z_L}-\phi)\tan(\phi_{Zin}) - \sin(\phi_{Z_L}-\phi)\right]}\\
X_{21} &=  \frac{R|Z_{in}|}{\sin(\phi_{Zin}+\phi) -\cos(\phi_{Zin}+\phi)\tan(\phi_{Z_L}) }\\
X_{22} &= X_{21} \frac{|Z_{L}|}{R|Z_{in}|} \frac{\cos(\phi_{Zin}+\phi)}{\cos(\phi_{Z_L})}
\end{align}

The ABCD matrix of the unit cell shown in Fig. \ref{fig:trl_circuit} is calculated by cascading the ABCD matrices of the sheet impedances and dielectric spacers. 

\begin{equation}
\begin{bmatrix}
A & B\\C & D
\end{bmatrix}=
\begin{bmatrix}
1 & 0\\1/Z_{s1} & 1
\end{bmatrix}\begin{bmatrix}
\cos(\beta d) & j Z_0\sin(\beta d) \\
j\sin(\beta d)/Z_0 & cos(\beta d)
\end{bmatrix}\begin{bmatrix}
1 & 0\\1/Z_{s2} & 1
\end{bmatrix}\begin{bmatrix}
\cos(\beta d) & j Z_0\sin(\beta d) \\
j\sin(\beta d)/Z_0 & cos(\beta d)
\end{bmatrix}\begin{bmatrix}
1 & 0\\1/Z_{s3} & 1
\end{bmatrix}
\label{eq:tl_abcd}
\end{equation}

Here, the sheet impedance values are $Z_{s1}$, $Z_{s2}$, and $Z_{s3}$, the characteristic wave impedance of the dielectric spacers is $Z_0$, and the propagation constant of the spacers is $\beta$.

The ABCD matrix of the desired unit cell (derived from the impedance matrix) is equated to the ABCD matrix given by \ref{eq:tl_abcd}, and the sheet impedance values are solved for.
\begin{align}
Z_{s1} &= \frac{jZ_0\sin(2\beta d)-\frac{(Z_0)^2}{Z_{s2}}\sin^2(\beta d)}{\frac{Z_{22}}{Z_{21}}-1+2\sin^2(\beta d) - j\frac{Z_0 \sin(2\beta d)}{2Z_{s2}}}\\
Z_{s2} &= \frac{(Z_0)^2 \sin^2(\beta d)}{jZ_0 \sin(2\beta d) - \frac{|Z|}{Z_{21}}} \\
Z_{s3} &= \frac{jZ_0\sin(2\beta d)-\frac{(Z_0)^2}{Z_{s2}}\sin^2(\beta d)}{\frac{Z_{11}}{Z_{21}}-1+2\sin^2(\beta d) - j\frac{Z_0 \sin(2\beta d)}{2Z_{s2}}}
\end{align}
Here, $|Z|$ is the determinant of the unit cell impedance matrix.

As a result, each sheet impedance is a function of the tangential field wave impedance on either side of the metasurface, the phase discontinuity across the metasurface, and the electrical length between the impedance sheets $(\beta d)$.

\section{Dolph-Tchebyscheff Far-Field Pattern Example}

In this section, we elaborate on the design of the compound metaoptic which produces the Dolph-Tchebyscheff far-field pattern. As described in the main text, the desired Dolph-Tchebyscheff pattern will have a main lobe at an azimuthal angle of 40 degrees and a maximum sidelobe level of -15 dB. The two constitutive metasurfaces are separated by a distance of $L=1.25\lambda$.

Dolph-Tchebyscheff far-field patterns are usually produced by linear phased arrays, where the complex-valued current of each array element is calculated using array theory \cite{rad_systems}. In this example, we calculated the current weights that produce the desired far-field pattern for an array of length $8\lambda$ with element spacing of $d_{elt}=\lambda/2$. Sinc function interpolation was used to obtain the desired continuous electric field distribution from the discrete current weights. In the sinc interpolation, the argument of the sinc function is set such that each interpolating sinc function has a maximum at the contributing element location and zeros at the neighboring element locations. Equation \eqref{eq:sinc_eq} is used to perform the sinc interpolation for the 17 current weights $(I_n)$ of the array.

\begin{equation}
E_{des} = \sum_{n=-8}^8 I_n\text{sinc}\left( \pi\frac{y-nd_{elt}}{d_{elt}}\right)
\label{eq:sinc_eq}
\end{equation}

\begin{figure}[ht]
\includegraphics[height=2.25in]{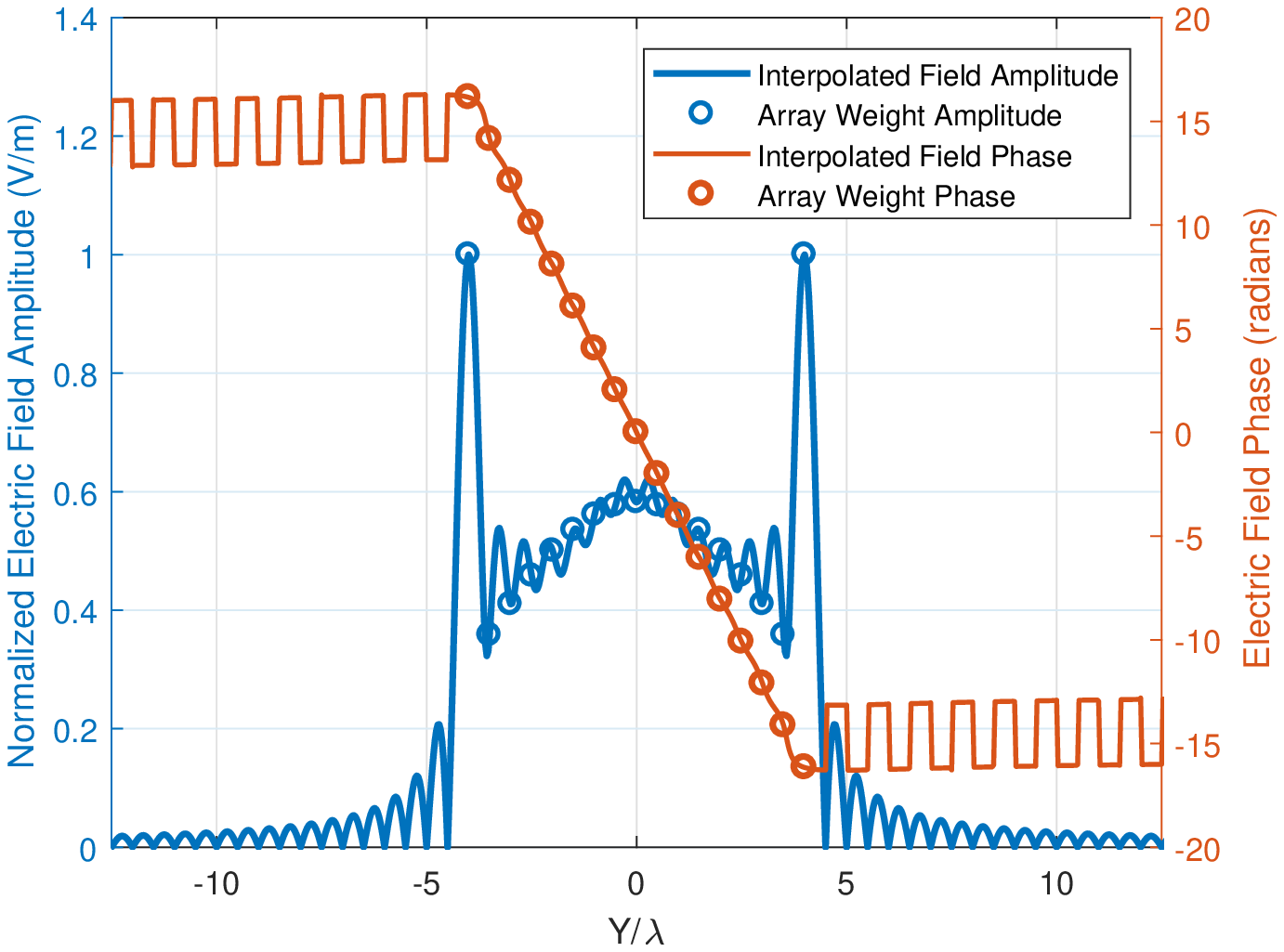}
\caption{Desired continuous electric field interpolated from the Dolph-Tchebyscheff element weights.}\label{fig:Edes_ideal}
\end{figure}

The desired electric field pattern is shown in Fig. \ref{fig:Edes_ideal}

\begin{figure}[ht]
\subfloat[\label{fig:s_GS_results}]{
	\includegraphics[width=0.32\textwidth]{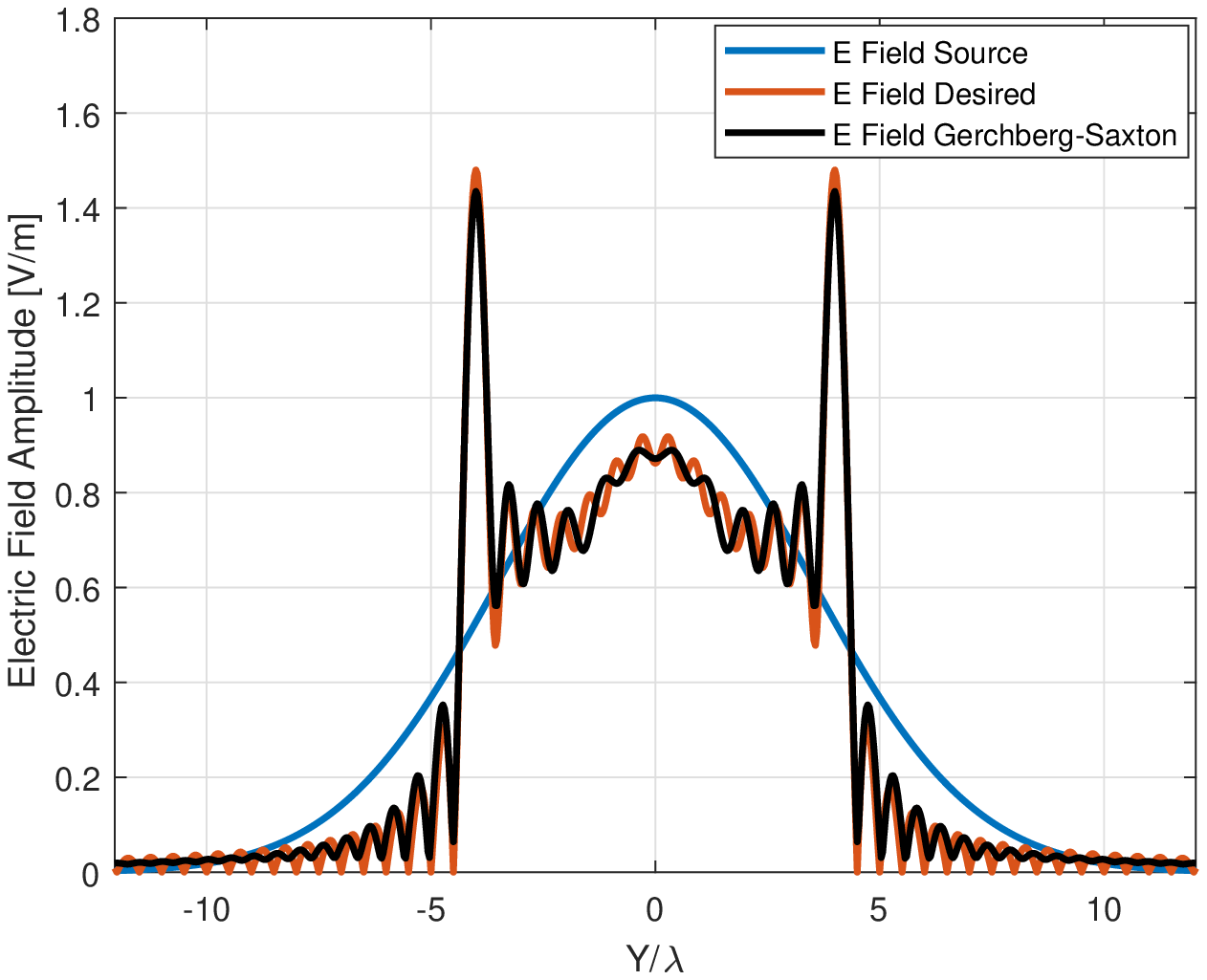}
	}
\subfloat[\label{fig:s_GS_PWspect}]{
	\includegraphics[width=0.32\textwidth]{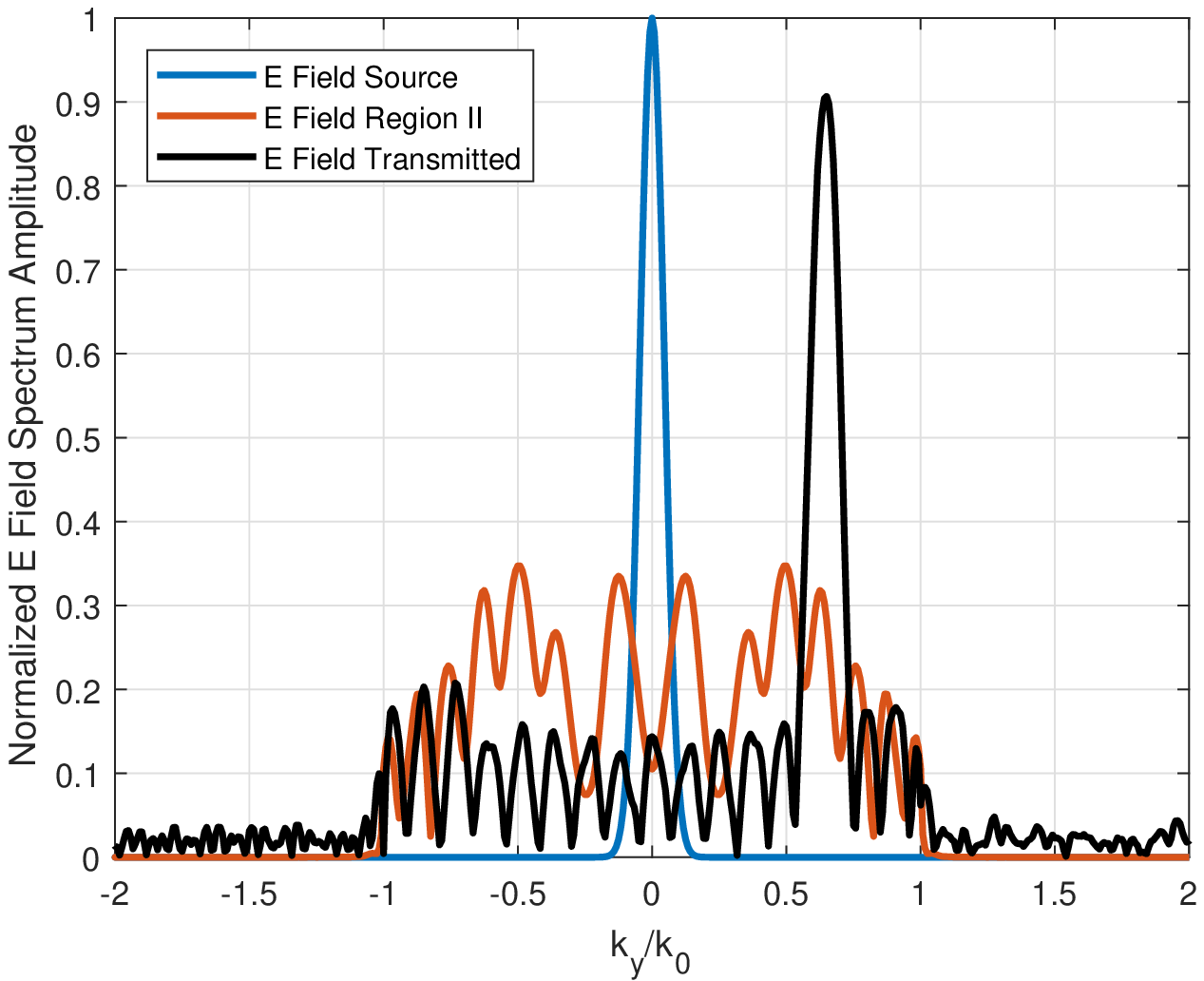}
	}
\subfloat[\label{fig:phase_discontinutiesMS}]{
	\includegraphics[width=0.32\textwidth]{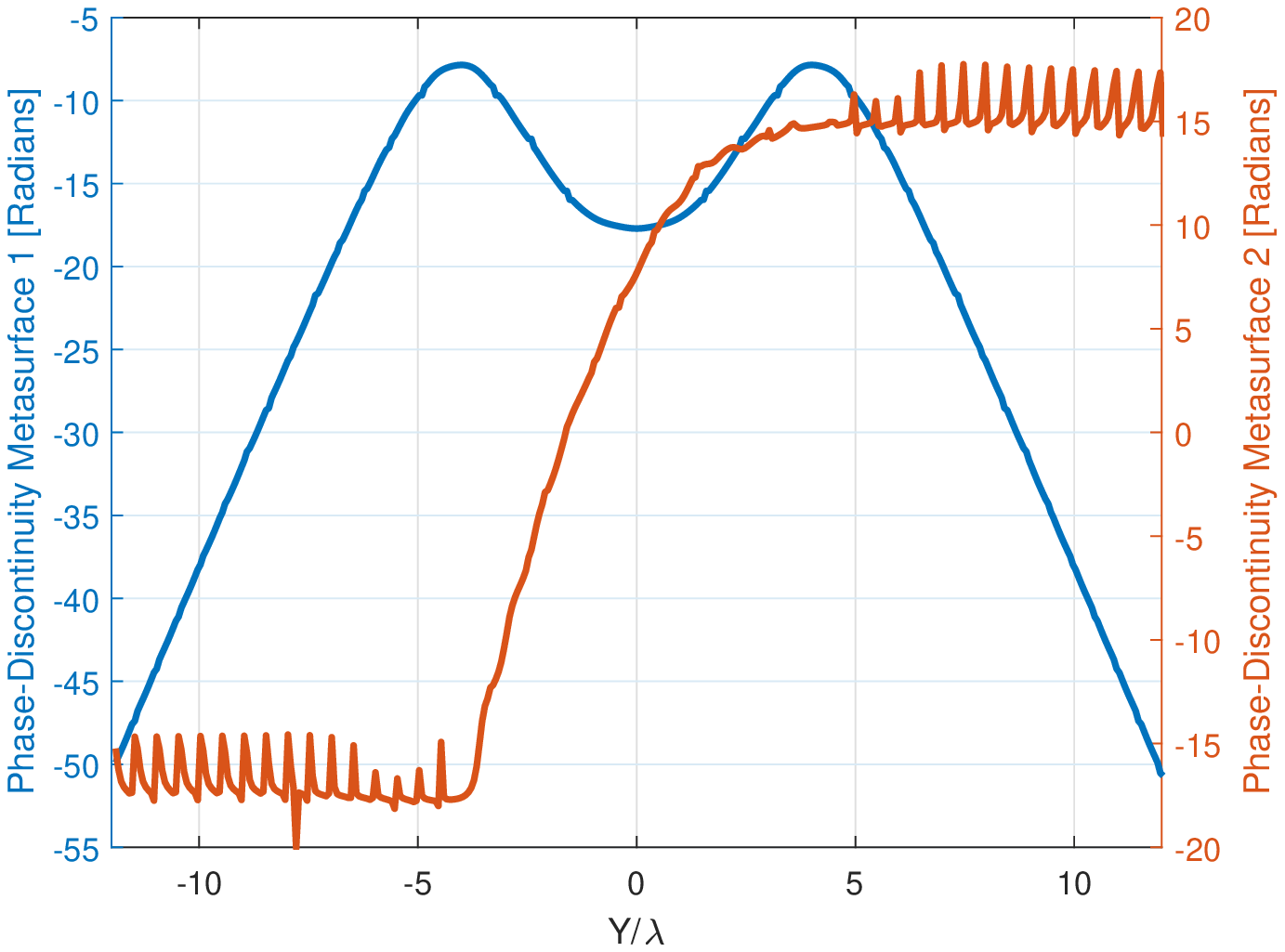}
	}
\caption{The plot in (a) shows the amplitude patterns for the source and desired field distributions, along with the field pattern resulting from the Gerchberg-Saxton procedure. The plot in (b) shows the plane wave spectrum magnitude before and after each metasurface in the metaoptic. The plot in (c) shows the calculated phase-discontinuty profile of each metasurface to form the Dolph-Tchebyscheff far-field pattern.}
\end{figure}

After setting the source and desired fields, the modified Gerchberg-Saxton phase retrieval algorithm is performed to determine the phase discontinuities of each metasurface. The phase-discontinuity of the first metasurface produces the desired amplitude pattern after a $1.25\lambda$ propagation distance, while the phase-discontinuity of the second metasurface produces the desired phase profile. Fig. \ref{fig:s_GS_results} shows the electric field amplitude profiles of the Gaussian source, the desired field, and the field resulting from the modified Gerchberg-Saxton algorithm. We see that the result of the Gerchberg-Saxton algorithm is slightly different from the desired field. This is due to the absence of evanescent content which would be necessary to accurately produce the highly oscillatory portions of the desired field. 

Figure \ref{fig:s_GS_PWspect} displays the plane wave spectrum of the electric field before and after each metasurface. We see there are three distinct spectra, one for each region of the compound metaoptic. The field of the Gaussian source has a Gaussian spectrum and the transmitted field spectrum has the desired Dolph-Tchebyscheff appearance. The plane wave spectrum  of the field in region II (between the metasurfaces) is spread widely across the propagating spectrum. Generally, a field pattern with evanescent content will not be exactly produced at the desired plane due to attenuation. The shape of the plane wave spectrum between the metasurfaces is a function of the relative differences between the source and desired field patterns, as well as the metasurface separation distance. Generally, the plane wave spectrum becomes more paraxial (close to the spectrum of the incident field) as the propagation distance is increased or the desired field pattern is changed to be more similar to the source field pattern. 

The phase discontinuity profiles of each metasurface are then calculated as the phase difference between the tangential electric fields on either side of the metasurface. Fig. \ref{fig:phase_discontinutiesMS} shows the phase discontinuity profile of each metasurface.

Once the phase discontinuity distributions of the metasurfaces are determined, the conservation of normal power flow condition is imposed and the electric field amplitude distribution transmitted by each metasurface is calculated. The fields tangential to each metasurface are now defined such that two phase-discontinuous reflectionless metasurfaces can reshape the amplitude and phase distributions of the incident beam into the desired complex-valued field pattern. 

\begin{figure}[h]
\centering
\begin{tabular}[b]{c }
	\subfloat[\label{fig:s_sim_E2D}]{
	\includegraphics[width = 0.455\textwidth]{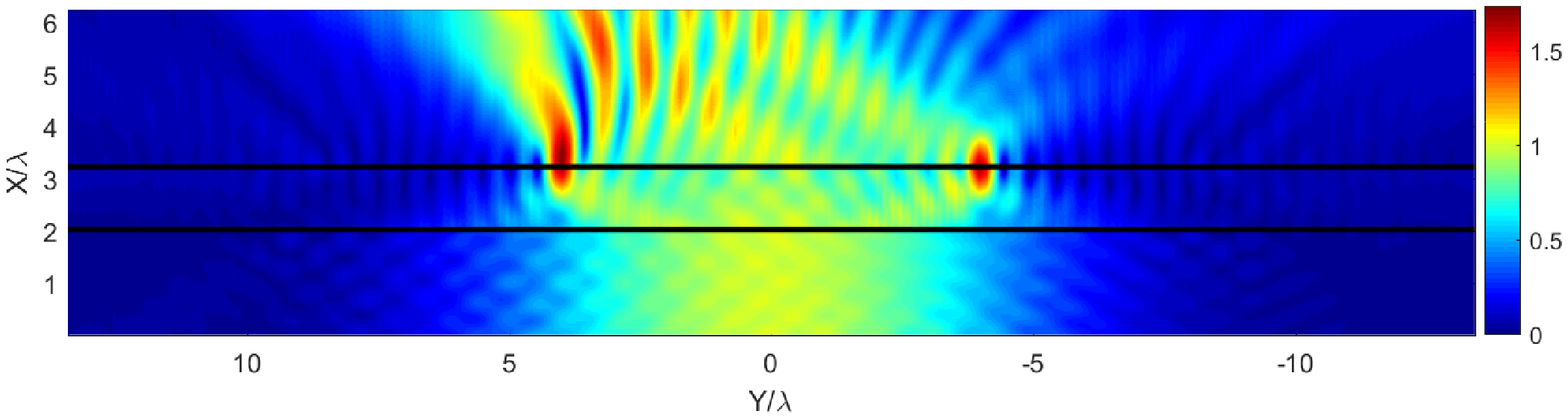}
    }\\
	 \subfloat[\label{fig:s_sim_E2D_real}]{
	\includegraphics[width = 0.455\textwidth]{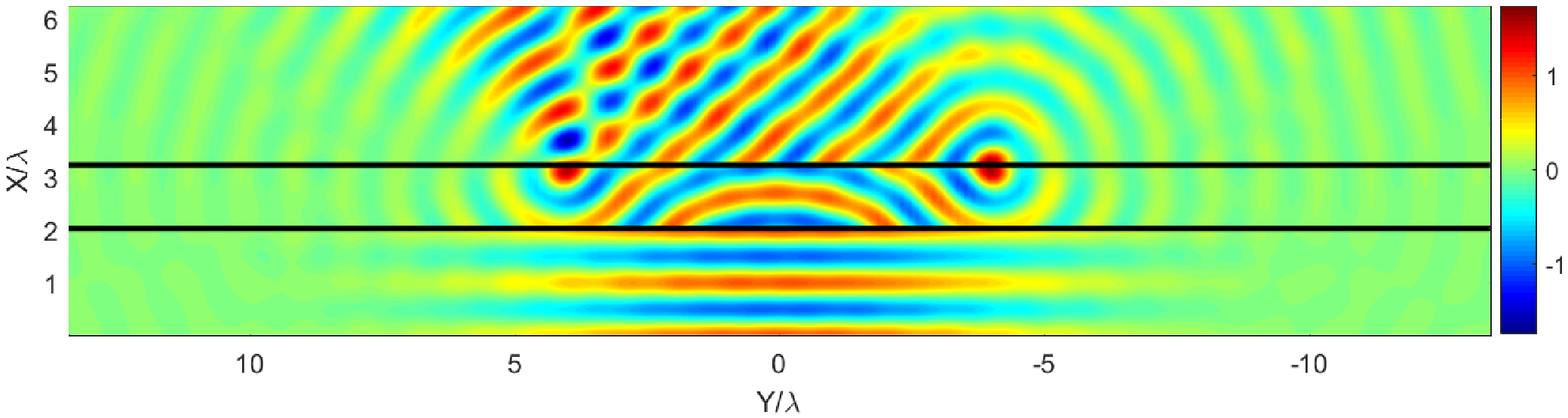}
    }   
\end{tabular}
\subfloat[\label{fig:s_sim_FF}]{
	\includegraphics[width = 0.415\textwidth]{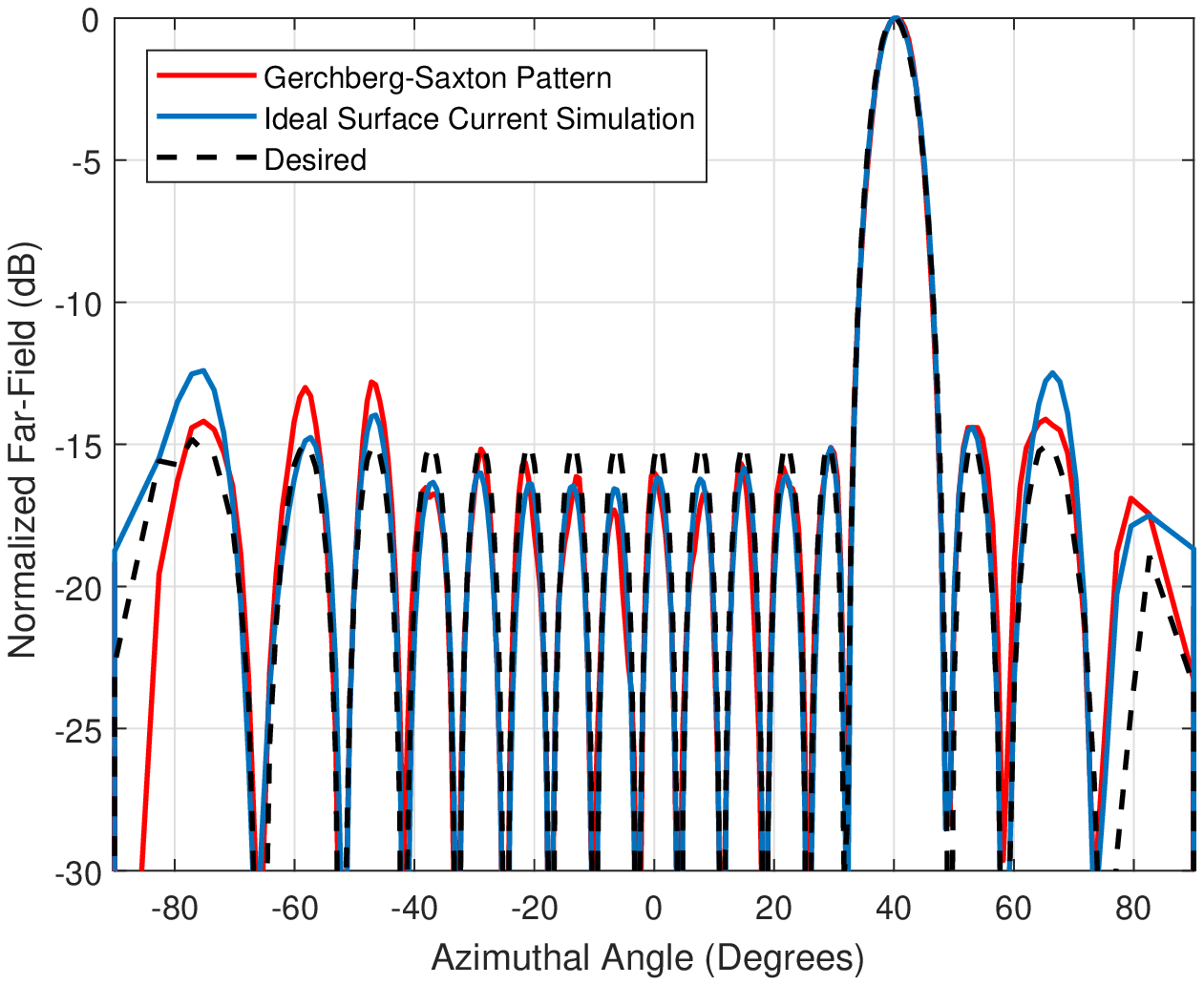}
    }\caption{The simulation results of the ideal surface current representation of each metasurface are shown in (a) as the complex magnitude of the electric field, in (b) as the real part of the electric field, and in (c) as the resulting far-field patterns.}
\end{figure}

The ideal performance of the compound metaoptic can be observed by replacing the metasurfaces with equivalent electric and magnetic surface current densities and performing a \emph{COMSOL Multiphysics} simulation. Fig. \ref{fig:s_sim_E2D} shows the magnitude of the complex electric field. Fig. \ref{fig:s_sim_E2D_real} shows the real part of the electric field. We see that the incident Gaussian beam is reshaped between the metasurfaces to form the amplitude pattern of the desired field, and with nearly no reflection. Fig. \ref{fig:s_sim_FF} shows the desired, Gerchberg-Saxton calculated, and simulated far-field patterns produced by the compound metaoptic. We see that the calculated and simulated far-field patterns closely match the desired Dolph-Tchebyscheff pattern, signifying that both the amplitude and phase of the Gaussian beam were altered by the metaoptic.

\begin{figure}[t]
\subfloat[\label{fig:s_ms1_zs}]{
	\includegraphics[height=2.25in]{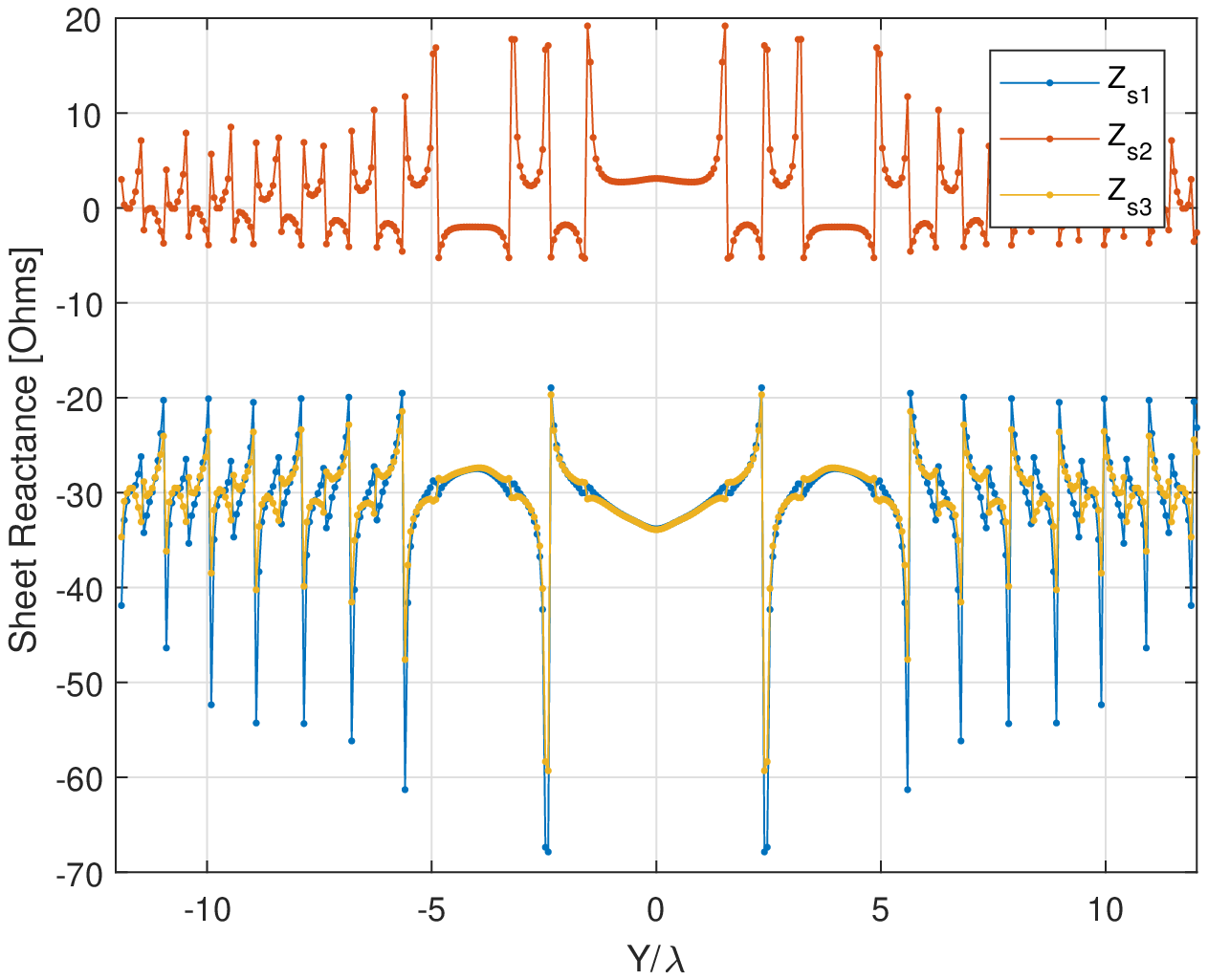}
	}
\subfloat[\label{fig:s_ms2_zs}]{
	\includegraphics[height=2.25in]{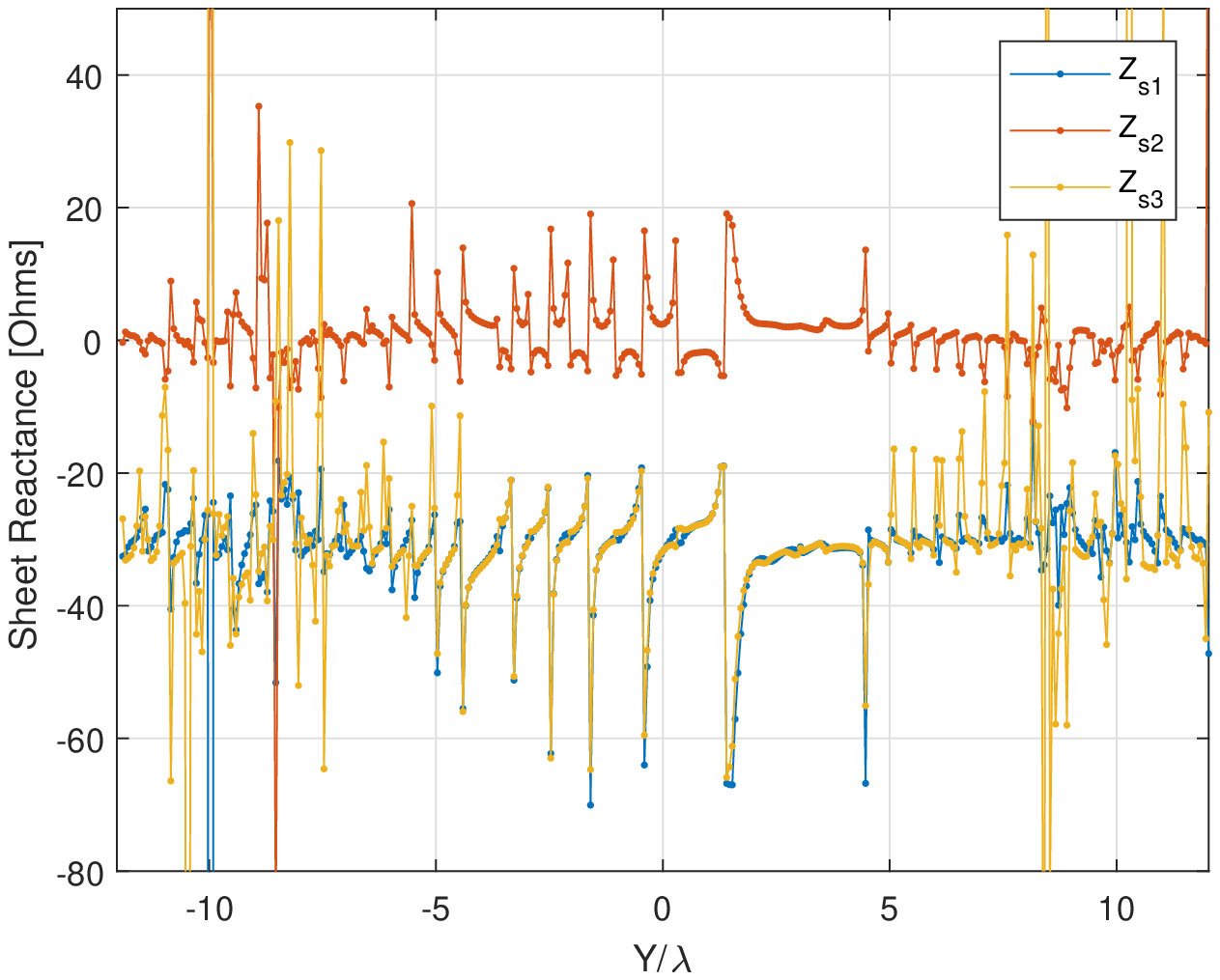}
	}
\caption{\label{fig:s_zs} The sheet impedance values to implement each bianisotropic metasurface of the compound metaoptic are shown for (a) the first metasurface and (b) the second metasurface.}
\end{figure}

The next step is to translate the parameters of each metasurface into the three sheet impedance values of the bianisotropic Huygens' metasurface unit cell (see Fig. 2 in the main text). The equations derived in Section \ref{sec:sheet_impedance} were used for this purpose. Fig. \ref{fig:s_zs} displays the three sheet impedance values for each unit cell along the metasurfaces. 

The electric sheet impedance values were then simulated in \emph{COMSOL} as ideal impedance boundaries, with the results included in the main text.

\section{Scattered field matching example}

In the second example, a compound metaoptic is designed to replicate the field scattered by three line scatterers. By radiating an electric field identical to the scattered field, the compound metaoptic produces a simple complex-valued hologram of the scatterers. In this example, three virtual line scatterers are placed beyond the metasurface at coordinates $(3\lambda, -2\lambda)$, $(1.5\lambda, 0.5\lambda)$, and $(4\lambda, 1.75\lambda)$ (see Fig. 4a of the main text). The origin is centered on the second metasurface of the compound metaoptic. The virtual scatterers are assumed to be illuminated by a plane wave traveling in the $-x$ direction. Therefore, the scatterers can be described as equal-magnitude line excitations with a phase dependent on the $x-$position of the scatterer.

Each line scatterer generates an outward-propagating cylindrical wave, taking the form of the zeroth order Hankel function of the second kind, $H_0^{(2)}(kr)$. The Fourier transform of this Hankel function is 
\begin{equation}
\mathscr{F}\{H_0^{(2)}(kr)\} = \frac{2}{\sqrt{k^2-k_y^2}}
\end{equation}
which is the plane wave spectrum of the scattered field for a line scatterer positioned at the origin. To calculate the plane wave spectrum of the complete scattered field (which is the desired field) in the $x=0$ plane, the plane wave spectrum of each scatterer is calculated relative to the origin and summed. 

\begin{equation}
\mathscr{E}_{des}(k_y) = \sum_{m=1}^3 \frac{2}{\sqrt{k^2-k_y^2}} \cdot e^{ik_y y_m }\cdot e^{ i k_x x_m} \cdot e^{i k x_m}
\label{eq:total_E_plane}
\end{equation}
Here, the first two exponentials serve to reverse propagate the electric field spectrum of each scatterer to the origin, where $(x_m, y_m)$ are the scatterer corrdinates, $k_x$ and $k_y$ are components of the plane wave spectrum, and $k$ is the wave number. The last exponential represents the phase of the illuminating plane wave.

\begin{figure}[!h]
\subfloat[\label{fig:Edes_spectrum_scatterers}]{
	\includegraphics[height=2.3in]{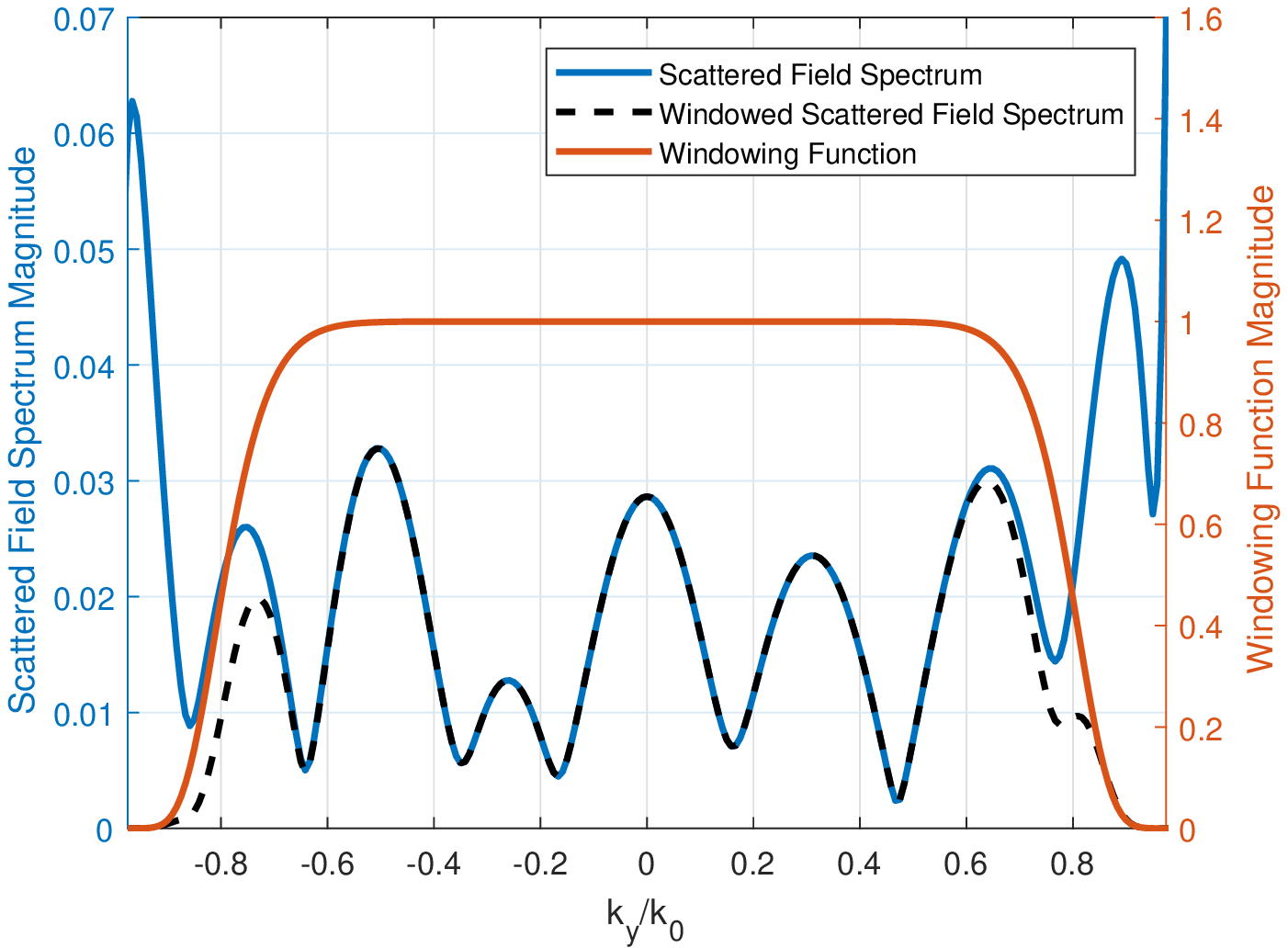}
	}
\subfloat[\label{fig:Edes_ideal_scatterers}]{
	\includegraphics[height=2.3in]{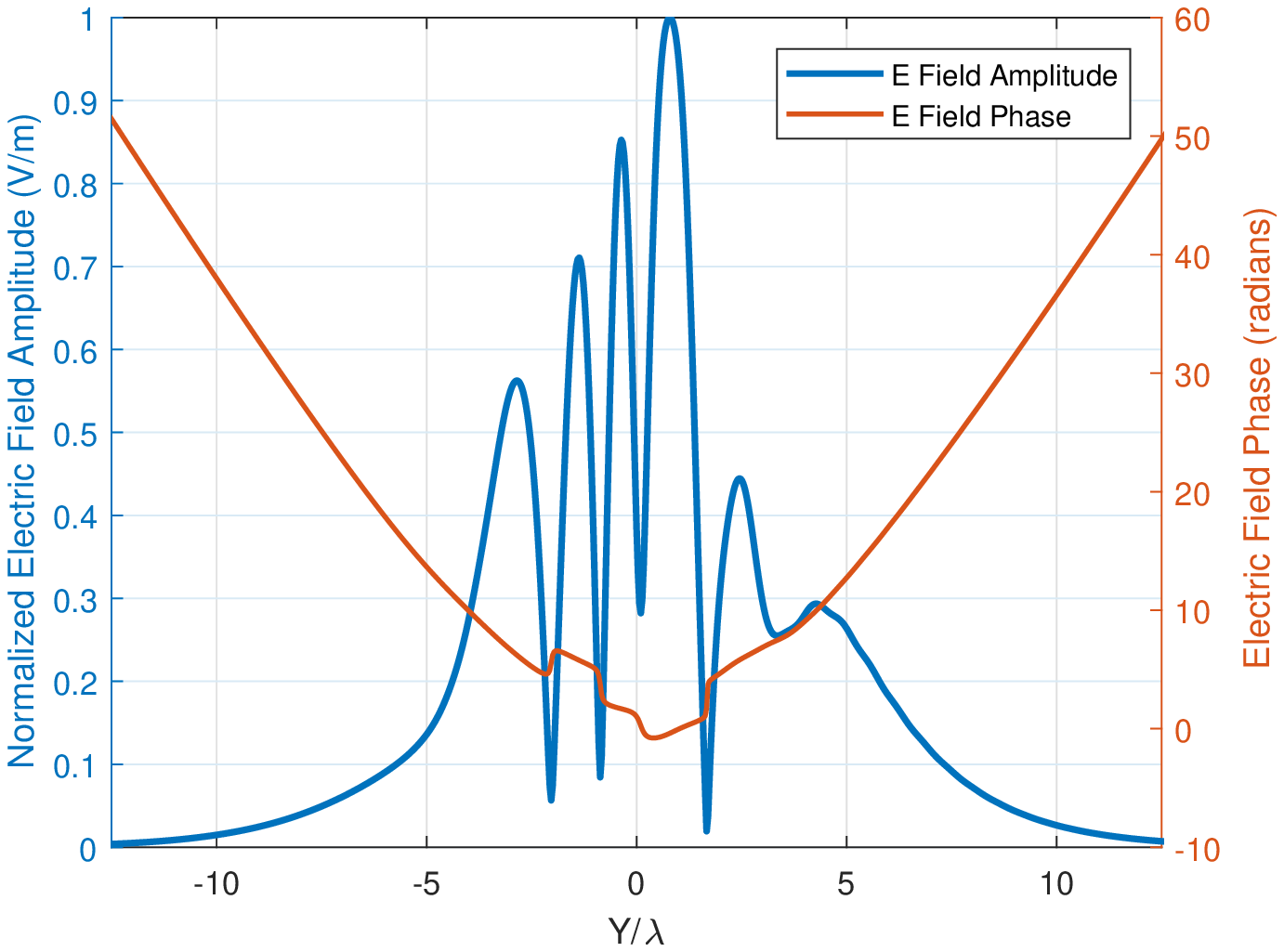}
	}
\caption{The plane wave spectra of the scattered field, the windowed scattered field, and the windowing function are shown in (a). The desired electric field in the spatial domain is shown in (b), which is produced from the windowed scattered field spectrum of the line scatterers.}\label{fig:Edes_scatterers}
\end{figure}

If the plane wave spectrum expressed in \eqref{eq:total_E_plane} is reproduced by the compound metaoptic, the transmitted field will be identical to that produced by the scatterers (a perfect hologram). However, since the line scatterers radiate isotropically, producing this field distribution would require an infinitely large aperture. To reduce the span of the required aperture, we apply a windowing function to the calculated plane wave spectrum. This action also limits the angular region over which the electric field produced by the compound metaoptic matches the desired scattered field. The window function used for this example is
\begin{equation}
f_{window}(k_y) = e^{ -12(k_y)^{14}/(3k_0)^{11.55}},
\end{equation}
which maintains a value greater than 0.96 for plane waves corresponding to $k_y > 0.64 |k_0|$ (equivalently, azimuthal angles between $\pm40$ degrees), and smoothly transitions to zero outside of this range. Fig. \ref{fig:Edes_spectrum_scatterers} displays a plot of the windowing function and the spectrum of the desired field. The desired electric field profile in the spatial domain is shown in Fig. \ref{fig:Edes_ideal_scatterers}. This is the field to be produced by the compound metaoptic. 

\begin{figure}[!h]
\subfloat[\label{fig:GS_result_scatterers}]{
	\includegraphics[height=2.3in]{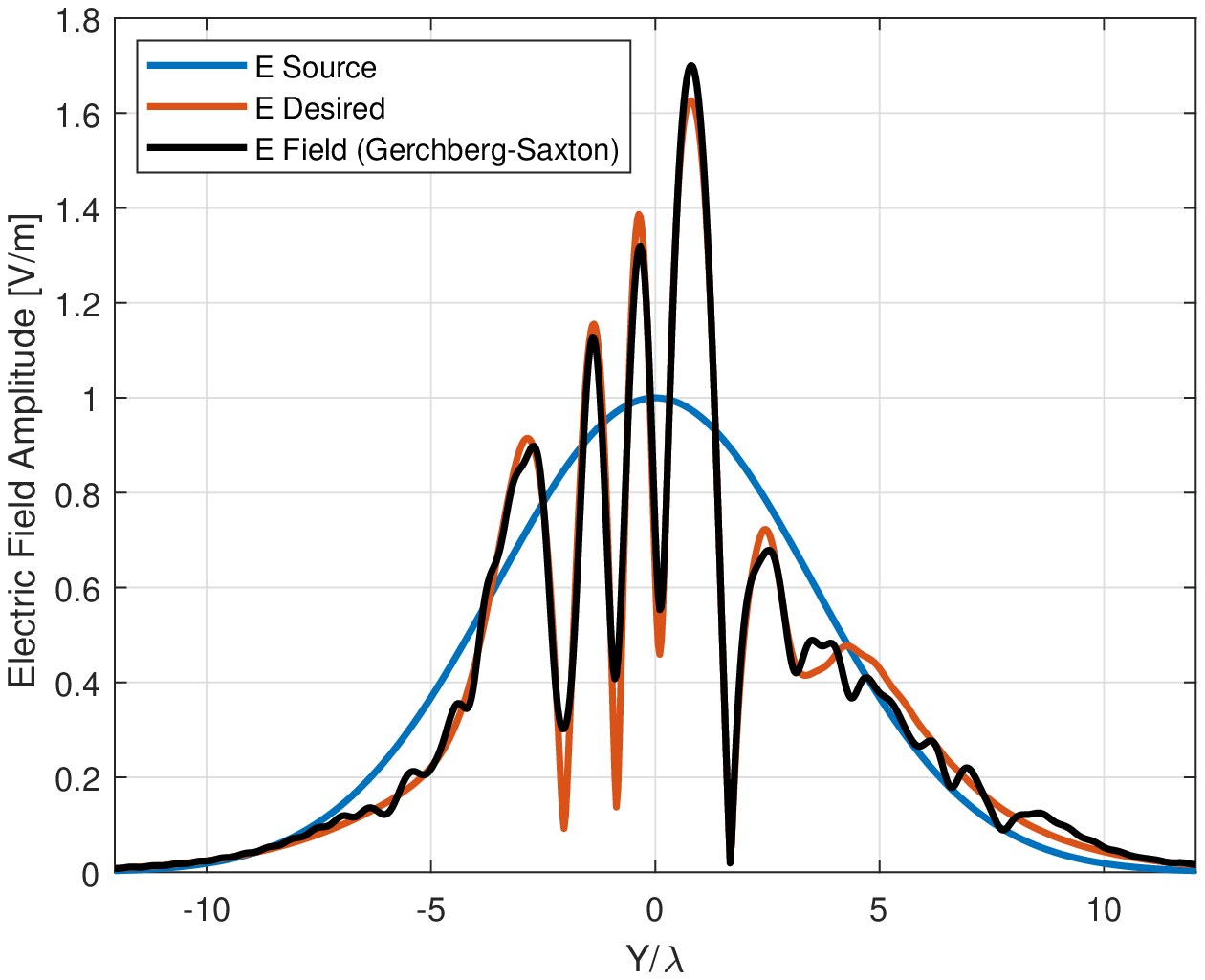}
	}
\subfloat[\label{fig:phase_disc_scatterers}]{
	\includegraphics[height=2.3in]{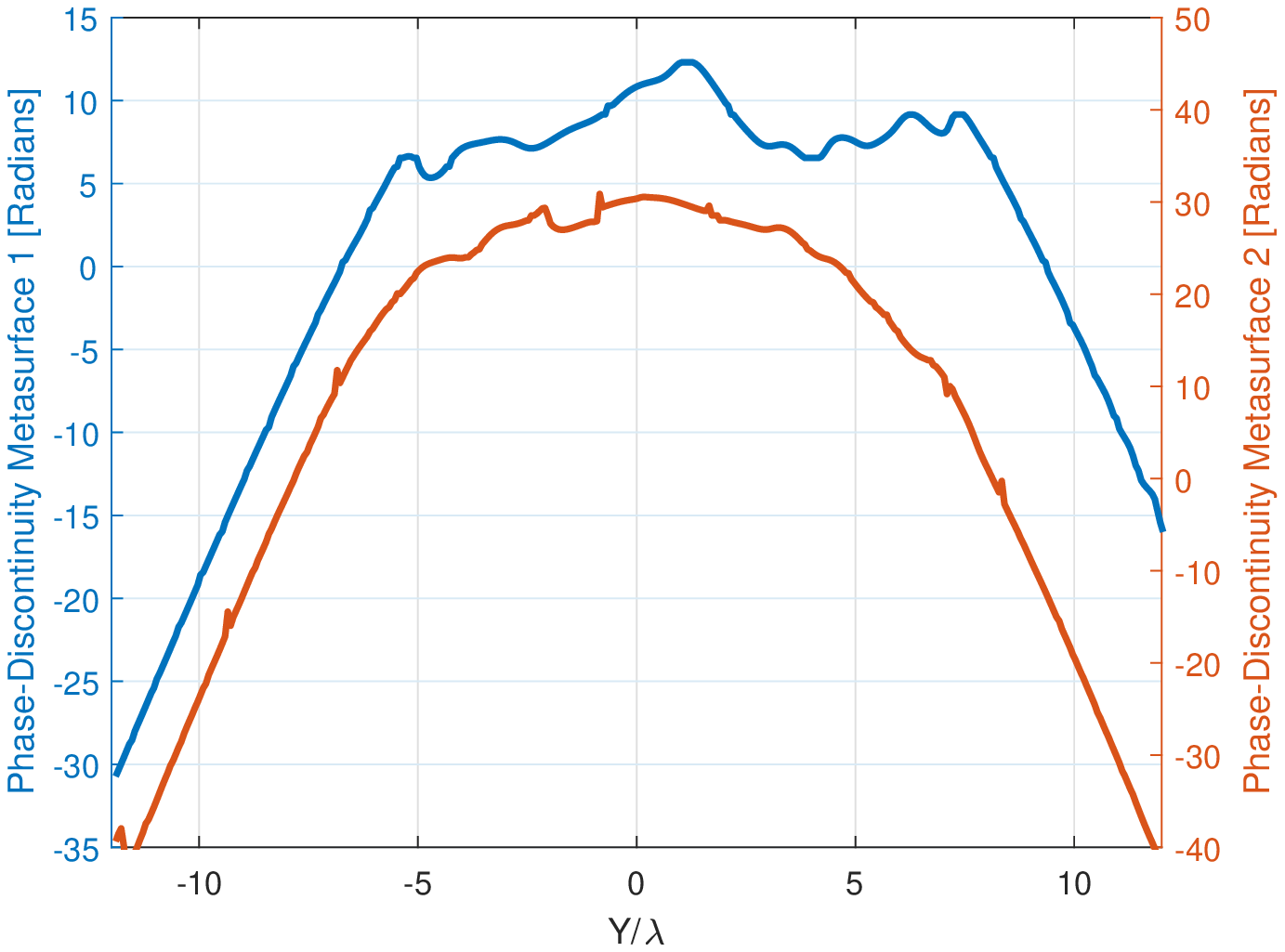}
	}
\caption{The results of the modified Gerchberg-Saxton algorithm. The plot in (a) shows the amplitude patterns for the source and desired field patterns, along with the field pattern resulting from the Gerchberg-Saxton procedure. The plot in (b) shows phase discontinuity of each metasurface. }
\end{figure}

Following the design procedure of the compound metaoptic, the modified Gerchberg-Saxton method is applied to determine the phase profile of each metasurface. Fig. \ref{fig:GS_result_scatterers} shows the electric field amplitude resulting from the Gerchberg-Saxton method, which closely matches the desired electric field amplitude distribution. The phase discontinuity profile of each metasurface is then calculated from the tangential fields and are shown in Fig. \ref{fig:phase_disc_scatterers}. Next, the electric field profiles are locally scaled to account for the conservation of normal power flow. This ensures zero reflection from either metasurface. 

\begin{figure}[t]
\subfloat[\label{fig:scatterers_ms1_zs}]{
	\includegraphics[height=2.25in]{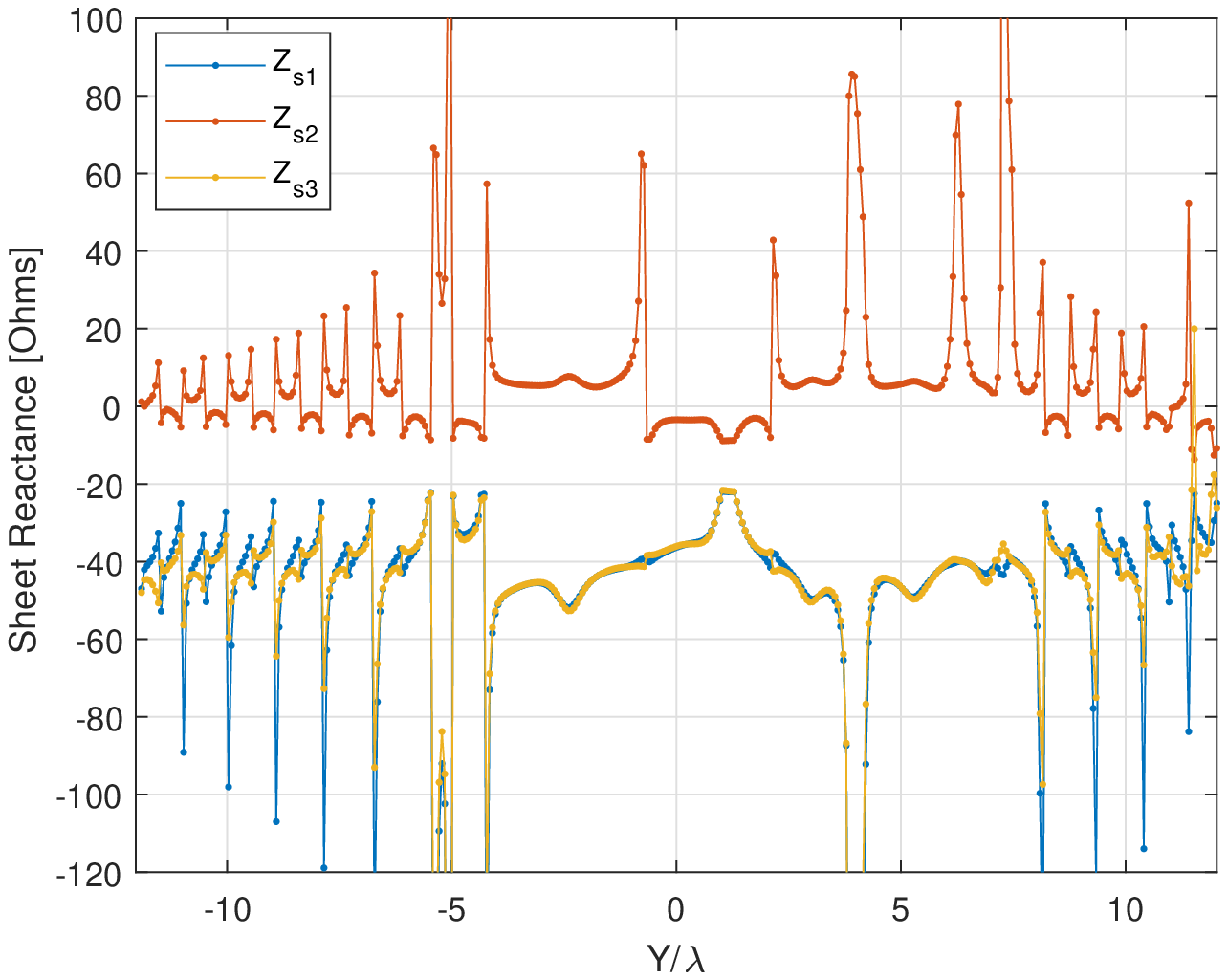}
	}
\subfloat[\label{fig:scatterers_ms2_zs}]{
	\includegraphics[height=2.25in]{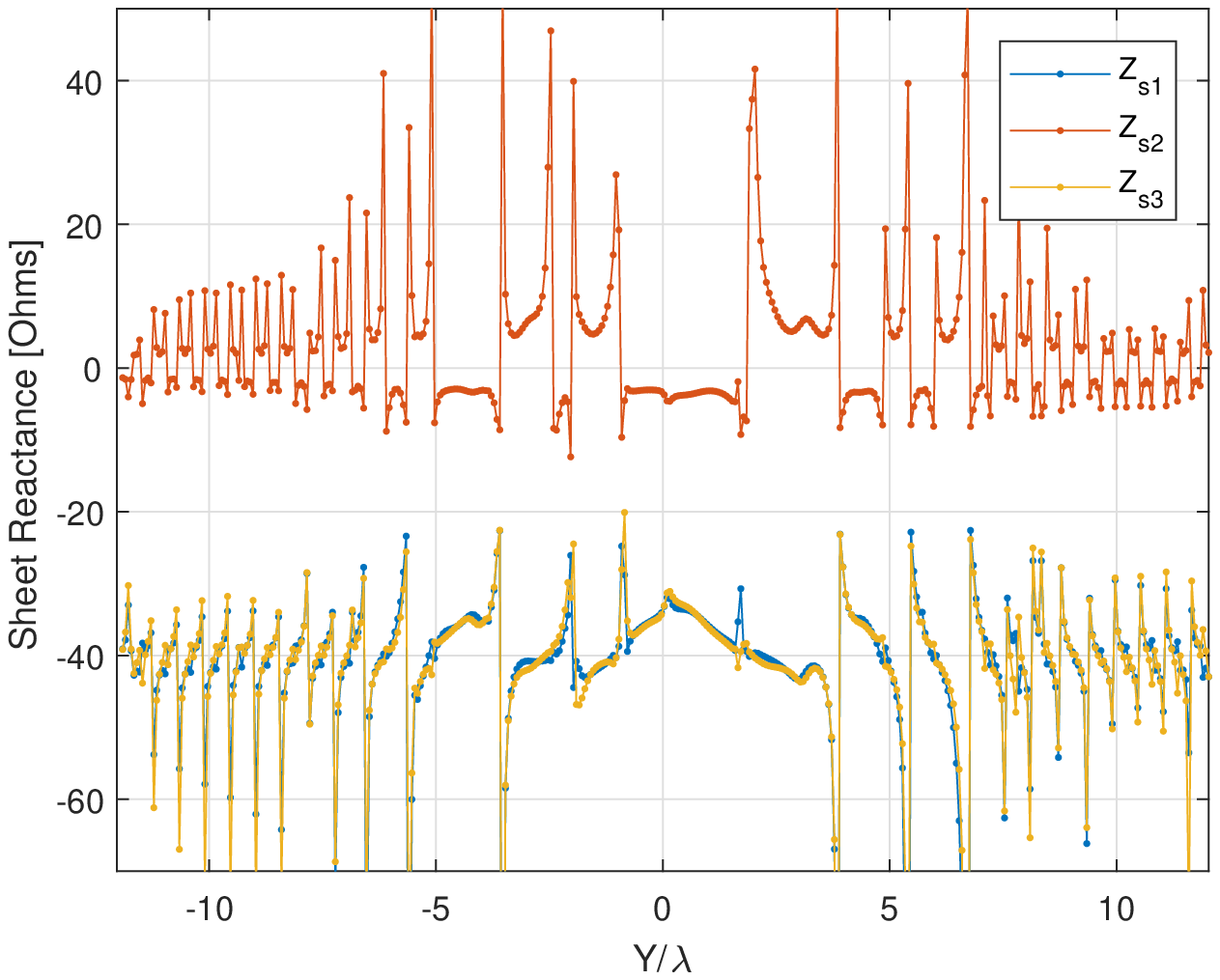}
	}
\caption{\label{fig:scatterers_zs} The sheet impedance values to implement each bianisotropic metasurface are shown for (a) the first metasurface and (b) the second metasurface.}
\end{figure}

Finally, with the fields tangential to each metasurface fully defined, each metasurface is implemented as a bianisotropic Huygens' metasurface. The three impedance sheet values for each metasurface unit cell are calculated and are plotted in Fig. \ref{fig:scatterers_zs}.

The electric sheet impedance values were then simulated in \emph{COMSOL} as ideal impedance boundaries, with the results reported in the main text.

\bibliography{supplementary_raeker_compound_metaoptic}